\documentclass[twocolumn]{aastex631}

\usepackage{amsmath}
\usepackage{multirow}
\usepackage{CJKutf8}

\received{September 20, 2021}
\revised{Feb 25, 2022}
\accepted{March 14, 2022}

\submitjournal{AJ}

\shorttitle{New IR Criterion for AGN to Lower Lumosities}
\shortauthors{Hviding et al.}

\graphicspath{{figures/}}

\begin{document}

\title{A New Infrared Criterion for Selecting Active Galactic Nuclei to Lower Luminosities}

\correspondingauthor{Raphael E. Hviding}
\email{rehviding@email.arizona.edu}

\author[0000-0002-4684-9005]{Raphael E. Hviding}
\affiliation{Steward Observatory, University of Arizona, 933 North Cherry Avenue, Tucson, AZ 85721, USA}

\author[0000-0003-4565-8239]{Kevin N. Hainline}
\affiliation{Steward Observatory, University of Arizona, 933 North Cherry Avenue, Tucson, AZ 85721, USA}

\author[0000-0002-7893-6170]{Marcia Rieke}
\affiliation{Steward Observatory, University of Arizona, 933 North Cherry Avenue, Tucson, AZ 85721, USA}

\author[0000-0002-0000-2394]{St\'ephanie Juneau}
\affiliation{NSF's National Optical-Infrared Astronomy Research Laboratory, 950 North Cherry Ave, Tucson, AZ 85719, USA}

\author[0000-0002-6221-1829]{Jianwei Lyu (\begin{CJK}{UTF8}{gbsn}吕建伟\end{CJK})}
\affiliation{Steward Observatory, University of Arizona, 933 North Cherry Avenue, Tucson, AZ 85721, USA}

\author[0000-0002-4940-3009]{Ragadeepika Pucha}
\affiliation{Steward Observatory, University of Arizona, 933 North Cherry Avenue, Tucson, AZ 85721, USA}

\begin{abstract}

    We present a spectroscopic and photometric analysis of a sample of 416,288 galaxies from the Sloan Digital Sky Survey (SDSS) matched to mid-infrared (mid-IR) data from the Wide-Field Infrared Survey Explorer (WISE). By using a new spectroscopic fitting package, GELATO (Galaxy/AGN Emission Line Analysis TOol), we are able to retrieve emission line fluxes and uncertainties for SDSS spectra and robustly determine the presence of broad lines and outflowing components, enabling us to investigate WISE color space as a function of optical spectroscopic properties. In addition, we pursue SED template fitting to assess the relative AGN contribution and nuclear obscuration to compare to existing mid-IR selection criteria with WISE. We present a selection criterion in mid-IR color space to select Active Galactic Nuclei (AGNs) with a $\sim$80\% accuracy and a completeness of $\sim$16\%. This is the first mid-IR color selection defined by solely using the distribution of Type I and Type II optical spectroscopic AGNs in WISE mid-IR color space. Our selection is an improvement of $\sim$50\% in the completeness of targeting spectroscopic AGNs with WISE down to an SDSS $r<17.77$\,mag. In addition, our new criterion targets a less luminous population of AGNs, with on average lower [\ion{O}{3}] luminosities by $\sim$30\% ($>0.1$\,dex) compared to typical WISE color-color selections. With upcoming large photometric surveys without corresponding spectroscopy, our method presents a way to select larger populations of AGNs at lower AGN luminosities and higher nuclear obscuration levels than traditional mid-IR color selections. 
    
\end{abstract}

\keywords{Active Galactic Nuclei (16), Active Galaxies (17), AGN Host Galaxies (2017)}

\section{Introduction}\label{sec:intro}

Active Galactic Nuclei (AGNs), objects powered by luminous accretion onto supermassive black holes (SMBHs) at the centers of galaxies, play a critical role in the evolution and growth of galaxies. AGN activity can heat and ionize gas in its host galaxy, potentially influencing star formation and inducing powerful feedback which can remove dust and gas \citep[e.g][and references therein]{fabianObservationalEvidenceActive2012,alexanderWhatDrivesGrowth2012}. In addition, AGN activity is a tracer of SMBH accretion and growth and is therefore vital for understanding the coevolution of host galaxies and their central SMBHs \citep{hopkinsCosmologicalFrameworkCoEvolution2008}. To assess of the contribution of AGNs to the growth of galaxies over cosmic time, it is essential to obtain a complete census of AGNs.

AGNs can be selected across a wide range of wavelengths as the AGN spectral energy distribution (SED) has signatures over the entire electromagnetic spectrum \citep[for a recent review, see][and references therein]{padovaniActiveGalacticNuclei2017}. However, different AGN selection techniques do not necessarily target the same objects. Critically, various selection techniques are sensitive to different AGN luminosities as compared to their host galaxies, and/or different levels of obscuration. In addition, AGN selected through distinct techniques have varying host galaxy properties, including galaxy colors, specific star formation rates, and clustering signals \citep[see][]{hickoxHostGalaxiesClustering2009,juneauWidespreadHiddenActive2013,ellisonStarFormationRates2016}.

A prominent spectroscopic selection method for AGNs relies on comparing the strength of optical and ultraviolet (UV) nebular emission lines. Historically, optical spectroscopy has been used to separate AGNs into two categories: Type I AGNs, characterized by blue UV and optical colors, high ionization narrow emission lines, and broad emission lines in optical/UV spectroscopy, and Type II AGNs, characterized by high ionization narrow emission lines and a lack of broad emission lines. In addition, a widely used set of diagnostics which compares the strength of Balmer lines to forbidden nebular transitions are the Baldwin, Phillips, and Terlevich diagrams \citep[BPT;][]{baldwinClassificationParametersEmissionline1981}. By comparing the narrow line component fluxes of the collisionally excited transitions to nearby Balmer lines, e.g. [\ion{O}{3}]$\lambda$5007\AA\ to H$\beta$, [\ion{N}{2}]$\lambda$6583\AA\ or [\ion{S}{2}]$\lambda$6717\AA\ or [\ion{O}{1}]$\lambda$6300\AA\ to H$\alpha$, the relative hardness of the ionization field can be determined. Throughout this work we will only make use of the [\ion{N}{2}] BPT diagram and will refer to it as the BPT diagram or diagnostic for simplicity.

Due to tight correlations between ionization parameter and metallicity, star-forming galaxies can be separated from active galaxies on the BPT diagram using the demarcation outlined in \citet{kauffmannHostGalaxiesActive2003}. In addition, \citet{kewleyTheoreticalModelingStarburst2001} used starburst and photoionization modeling to determined a regime of the BPT diagram where the line ratio strengths cannot be explained by star formation alone, resulting in demarcations above which objects are considered to have their emission lines predominately driven by AGN activity. Those objects which are found above the \citet{kauffmannHostGalaxiesActive2003} demarcation but below the \citet{kewleyTheoreticalModelingStarburst2001} demarcation are historically said to be `composite' galaxies, where both AGN and star-forming processes may contribute to the emission line ratios.

In the mid infrared (mid-IR), AGN selection techniques rely on targeting emission from hot dust in the circumnuclear torus. 
The mid-IR is of particular interest in searching for obscured AGNs, where the X-ray and UV-optical signatures are masked by intervening gas and dust respectively \citep{hickoxObscuredActiveGalactic2018}.
In 2009, the Wide-Field Infrared Survey Explorer \citep[WISE;][]{wrightWidefieldInfraredSurvey2010}, a space-based mid-IR telescope, was launched. WISE observed the whole sky in four photometric bands, centered at 3.4, 4.6, 12, and 22\,$\mu$m, referred to as \textit{W1}, \textit{W2}, \textit{W3}, and \textit{W4} respectively.
In this work, we refer to the measured flux in a given photometric band as $f_{\rm{band}}$.
To probe for the characteristic AGN infrared power law, adjacent bands from WISE are compared to search for red colors. The hot dust emission from the torus heated by AGN activity will result in high flux ratios of $f_{\rm \textit{W2}}/f_{\rm \textit{W1}}$ and $f_{\rm \textit{W3}}/f_{\rm \textit{W2}}$, i.e. red WISE \textit{W1$-$W2} and \textit{W2$-$W3} colors.

In this work, we primarily focus on selection of AGNs in the optical and mid-IR regimes to understand the distribution of Type I and Type II AGNs in mid-IR color space. It becomes ever more important to find ways to select AGNs through photometry alone due to the advent of large photometric surveys without corresponding spectroscopic observations from the next generation of optical and near-IR telescopes, e.g. Euclid \citep{laureijsEuclidDefinitionStudy2011}, Nancy Grace Roman Space Telescope \citep[formerly WFIRST]{spergelWideFieldInfrarRedSurvey2015}, SPHEREx \citep{doreScienceImpactsSPHEREx2016}, and Vera C. Rubin Observatory \citep[formerly LSST]{ivezicLSSTScienceDrivers2019}.

Due to the availability of WISE data over the entire sky, many authors have assembled large samples of infrared-selected AGNs. \citet{jarrettSpitzerWISESurveyEcliptic2011} synthesized WISE colors of AGNs by combining templates of galaxies observed by the Spitzer Space Telescope along with modeling to separate AGNs from star-forming galaxies, resulting in a selection box.
\citet{sternMidInfraredSelectionAGN2012} made use of empirical galaxy templates from \citet{assefLowResolutionSpectralTemplates2010} to derive a single \textit{W1$-$W2} color criterion for selecting on AGN activity for objects with $\textrm{\textit{W2}}<15$\,mag.
This was later expanded upon in \citet{assefMIDINFRAREDSELECTIONACTIVE2013} which used SED modeling to push to fainter magnitudes with a \textit{W1$-$W2} versus \textit{W2} selection criterion, culminating in the 90\% reliability (R90) and 75\% completeness (C75) WISE AGN Catalogs with over 4 million and 20 million AGN candidates respectively \citep{assefWISEAGNCatalog2018}.
In addition, \citet{mateosUsingBrightUltraHard2012} identified a region of mid-IR color space that is occupied by AGNs selected using hard X-ray surveys and confirmed with optical spectroscopy. This mid-IR color region is referred to at the \citet{mateosUsingBrightUltraHard2012} wedge and was used by \citet{secrestIdentificationMillionActive2015} to identify 1.4 million AGNs over the entire sky.

As mid-IR selection is designed to probe for the existence of a power law from hot dust components, it is biased towards selecting AGNs whose output dominates over the host galaxy at mid-IR wavelengths. Therefore, AGNs exist in regions outside those targeted by \citet{jarrettSpitzerWISESurveyEcliptic2011}, \citet{sternMidInfraredSelectionAGN2012} \citet{mateosUsingBrightUltraHard2012}, \citet{assefMIDINFRAREDSELECTIONACTIVE2013} and other authors. Conversely, those regions believed to be dominated by AGNs may host significant fractions of star-forming galaxies with mid-IR colors masquerading as AGNs. \citet{lamassaSDSSIVEBOSSSpectroscopy2019} found that over 40\% (50\%) of objects satisfying the \citet{assefMIDINFRAREDSELECTIONACTIVE2013} R90 (R75) criterion are spectroscopically identified as star-forming, while \citet{mendezPRIMUSInfraredXRay2013} found star-forming galaxies contaminate the \citep{sternMidInfraredSelectionAGN2012} at specific redshifts. Conversely, \citet{hainlineSpectroscopicSurveyWISEselected2014} and \citet{hvidingCharacterizingWISEselectedHeavily2018} studied regions of mid-IR space outside the \citet{mateosUsingBrightUltraHard2012} wedge to find heavily obscured quasars missing in X-ray surveys of AGNs. While optical spectroscopy has been used in some cases to validate mid-IR AGN color selection criteria, there has not been a study that has defined a color selection using optical spectroscopic AGNs alone.

In this work, we undertake a systematic examination of WISE color space to investigate the optical spectroscopic properties of mid-IR selected AGNs and the distribution of spectroscopic AGNs in mid-IR color space. 
In Section \ref{sec:sample}, we assemble a large representative spectroscopic galaxy sample of 416,288 objects matched to mid-IR photometry from WISE. 
Critically, to identify the presence of broad components and other features in emission lines, we make use of a new spectroscopic fitting routine to robustly determine optical spectroscopic properties and the presence of additional features in ionized emission lines described in Section \ref{sec:spec}. In Section \ref{sec:sed}, we undertake template-fitting SED modeling to compare against similar endeavors in the literature which attempt to determine the presence of an AGN by examining the galaxy SED. Our results from combining the spectroscopic and photometric fitting are presented in Section \ref{sec:result}. We introduce a new selection criterion for optical identified objects matched to mid-IR photometry to select optical spectroscopic AGNs in WISE color-color space. Finally, we present our conclusions and discussion in Section \ref{sec:conc}.

Throughout this work we assume a flat Lambda Cold Dark Matter ($\Lambda$CDM) cosmology with H$_0=70$\,km\,s$^{-1}$\,Mpc$\,$ and $\Omega_{\rm m}=0.3$. Throughout this work we use AB magnitudes for optical SDSS photometry and Vega magnitudes for mid-IR WISE photometry.

\begin{figure*}[ht!]
    \includegraphics[width=\textwidth]{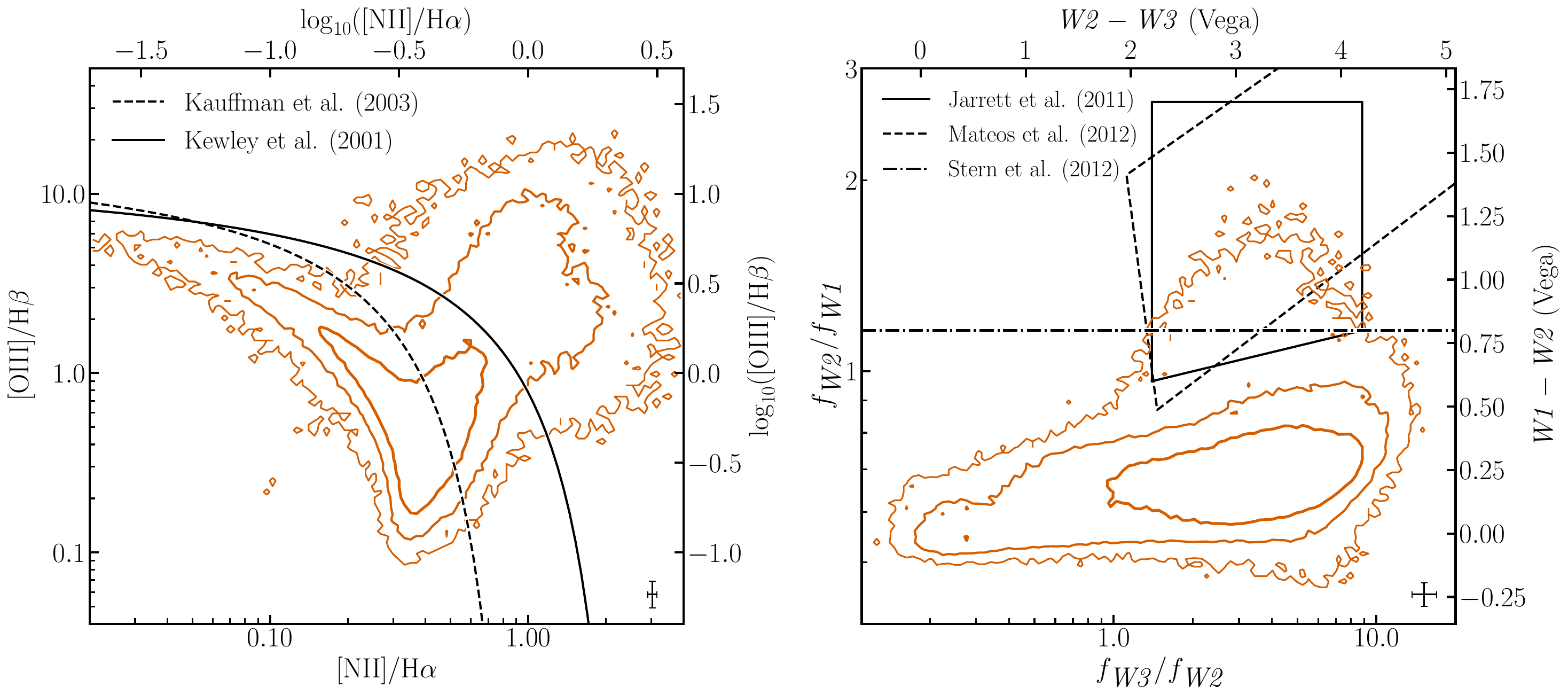}{}
    \caption{The WISE color-color and BPT diagnostic distributions of the WISE Matched SNR sample presented in this work. In the left panel we plot the BPT diagnostic for objects in our WISE Matched SNR sample with a flux SNR $>3$ detection in each of the BPT emission lines. The \citet{kauffmannHostGalaxiesActive2003} (dashed) and \citet{kewleyTheoreticalModelingStarburst2001} (solid) diagnostic curves are plotted as well. In the right panel, we plot the WISE color-color distribution of our WISE Matched SNR sample. Selected AGN color selection criteria are overplotted: the \citet{jarrettSpitzerWISESurveyEcliptic2011} box (solid), the \citet{mateosUsingBrightUltraHard2012} wedge (dashed), and the \citet{sternMidInfraredSelectionAGN2012} line (dot-dashed). In both panels we present orange contours encompassing 68\%, 95\%, and 99.7\% of the corresponding plotted sample and show error bars corresponding to the median uncertainty in the bottom right corner.\label{fig:sample}}
\end{figure*}

\section{Sample Selection \& Properties}\label{sec:sample}

Optical spectroscopy is well suited for probing the ionization state of the nebular gas in galaxies out to moderate redshifts ($z\lesssim 1$) through the measurement of strong nebular emission lines \citep[e.g.][]{trouilleOPTXProjectIdentifying2011,juneauNewDiagnosticActive2011}. To determine the presence and strength of AGN activity in a large sample of galaxies, we turn to the Sloan Digital Sky Survey \citep[SDSS;][]{yorkSloanDigitalSky2000}, a dedicated optical spectroscopic and photometric survey across the $u$, $g$, $r$, $i$, and $z$ bands \citep{gunnSloanDigitalSky1998} to assemble a representative spectroscopic galaxy sample. Using SDSS, samples of tens of thousands of spectroscopically identified AGNs have been constructed \citep[e.g.][]{kauffmannHostGalaxiesActive2003,kewleyHostGalaxiesClassification2006}.

We begin with the SDSS Main Galaxy Sample (MGS) from the Legacy subsurvey. This consists of a highly complete sample ($>99$\%) of astronomical objects with optical spectroscopy down to an $r$-band magnitude of $r\leq17.77$ and an $r$-band half-light surface brightness $\mu_{50} \leq 24.5$ in a 7,500\,deg$^2$ region in the northern hemisphere and a 740\,deg$^2$ region in the southern hemisphere \citep{straussSpectroscopicTargetSelection2002}. We start assembling our sample from the SDSS DR17 data release \citep{abdurroufSeventeenthDataRelease2021} using the CasJobs utility on the SDSS SkyServer system by querying all SDSS spectroscopic objects with the Legacy MGS flag. This results in 708,557 unique objects comprising a total of 770,697 spectra.

Given that many objects in the MGS have multiple spectra, we only select one spectroscopic result per object to avoid biasing our results. We select the spectroscopic result with the best fit as described in Section \ref{subsec:results}. Therefore, when we present statistics computed on the Legacy MGS or the samples defined in this work, we will not overrepresent objects with multiple spectroscopic observations.

For each spectrum, we make use of the automated SDSS redshift, which we denote as $z_0$, as an initial estimate of the redshift throughout this work. In addition, we denote $\sigma_{z_0}$ as the provided error on $z_0$. To remove objects for which there is no spectroscopic information we exclude all objects that have one or more of the following: (1) any \texttt{zWarning} flag (6,357 spectra, $0.82\%$), (2) a catastrophic redshift failure, either from an unplugged fiber or the automated pipeline, identified as having a $\sigma_{z_0} = 0$ or $=-1$ (2,682 spectra, $0.35\%$), (3) a low-confidence redshift, which we define as objects with $\sigma_{z_0}/(1+z) > 1$ (207 spectra, $0.03\%$). Following this process we are left with 764,229 spectra corresponding to 702,993 unique galaxies, $99.21\%$ of the SDSS MGS. We denote this as our `Spectroscopic' sample. As these spectra are removed primarily for reasons unrelated to the targets themselves we consider the Spectroscopic sample nearly as complete as the parent sample.

To investigate the mid-IR properties of our Spectroscopic sample, we match the SDSS Legacy MGS to WISE from the allWISE \citep{cutriVizieROnlineData2013} data release using the crossmatching utility at the NASA/IPAC Infrared Science Archive (IRSA). As SDSS positions are given in the reference frame consistent with the Fifth Fundemental Catalog \citep[FK5]{frickeFifthFundamentalCatalogue1988} J2000.0 epoch, and WISE observations use the International Celestial Reference Frame \citep[ICRF]{maInternationalCelestialReference1998}, we take the SDSS coordinates and transform them accordingly. We take the closest match within a 2\arcsec\ radius to achieve a false matching rate of less than 1\% \citep{krawczykMeanSpectralEnergy2013}. This results in our `WISE Crossmatched' sample with 674,631 unique objects comprising a total of 733,404 spectra.

To place objects on the typical WISE color-color diagram, we require a confidence in the represented fluxes, i.e. \textit{W1}, \textit{W2}, and \textit{W3}, at the 3$\sigma$ level equivalent to a likelihood of $\sim$99.87\%. Throughout this work we will use a 3$\sigma$ threshold to ensure consistent detections. For the WISE photometry we achieve this by requiring a flux signal-to-noise ratio (SNR) of $>3$ in each of the referenced bands. In doing so we reject 258,343 objects comprising 281,081 spectra which we refer to as our `WISE SNR Rejected' sample. Our WISE matched and SNR restricted sample is denoted as the `WISE Matched SNR' sample with 416,288 unique objects comprised of 452,323 spectra. Our WISE Matched SNR sample represents all SDSS Legacy MGS objects with confident WISE detections that have available spectroscopy and is the sample used for analysis throughout this work. The WISE Matched SNR sample will enable comparisons of the optical spectroscopic properties across WISE color-color space for a representative sample of galaxies.

We present the BPT diagnostic and WISE color-color distributions of the WISE Matched SNR sample in Figure \ref{fig:sample}. We plot the BPT diagram for galaxies where the BPT emission lines can be detected at the 3$\sigma$ level in Section \ref{sec:spec}. In addition, we plot the distribution of $f_{\rm \textit{W2}}/f_{\rm \textit{W1}}$ versus $f_{\rm \textit{W3}}/f_{\rm \textit{W2}}$ for the galaxies in the WISE Matched SNR sample along with the \citet{jarrettSpitzerWISESurveyEcliptic2011} box, \citet{sternMidInfraredSelectionAGN2012} line, and \citet{mateosUsingBrightUltraHard2012} wedge to illustrate typical AGN selection criteria in WISE color space.

\begin{figure*}[ht!]
    \includegraphics[width=\textwidth]{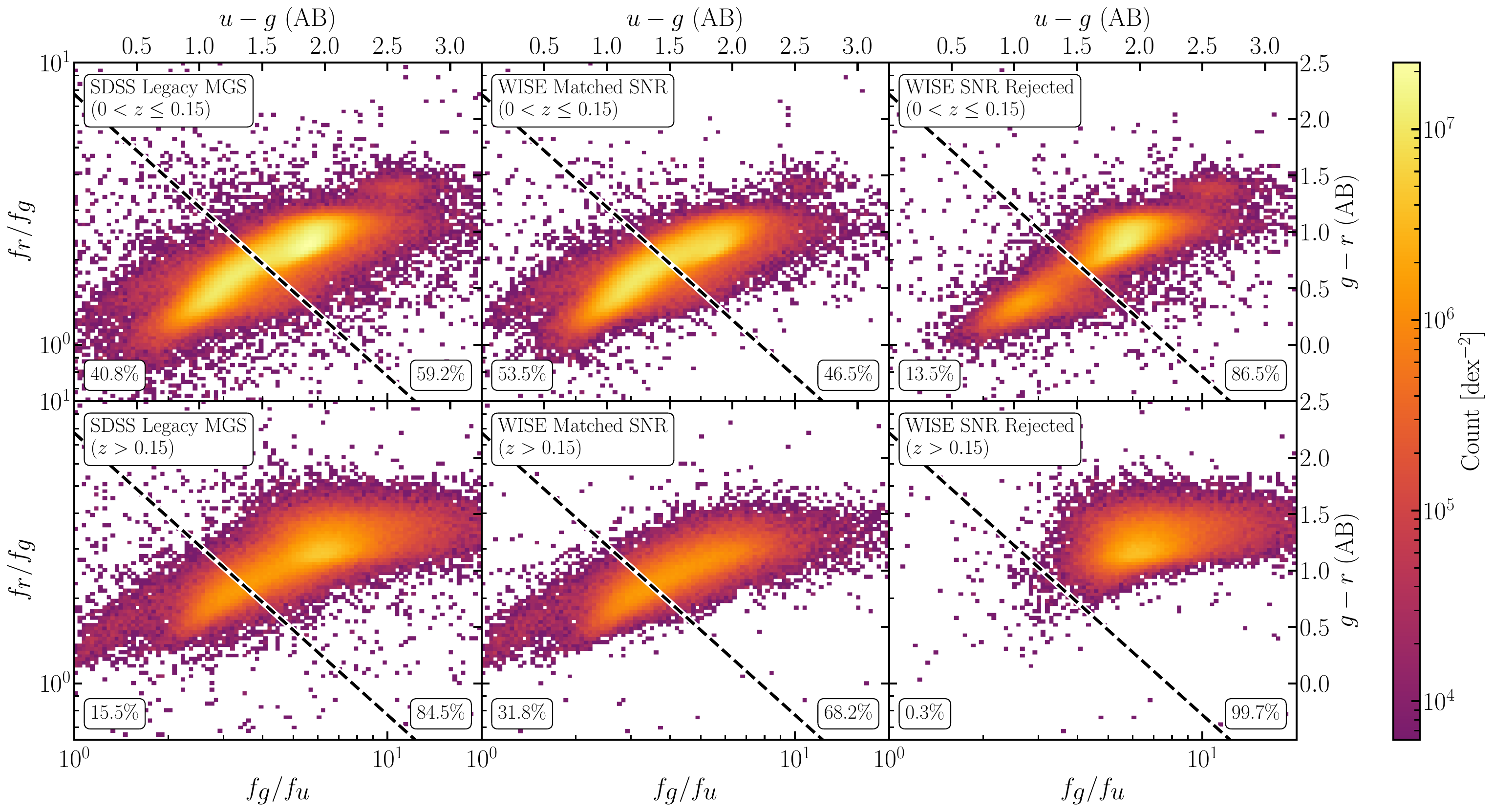}{}
    \caption{SDSS $f_r/f_g$ versus $f_g/f_u$ distributions of the SDSS Legacy MGS (left), our WISE Matched SNR sample (center), and our WISE SNR Rejected sample (right). In the top panels we plot the low-redshift regime ($0<z\leq0.15$) of the samples and on the bottom samples we show the high-redshift regime ($z > 0.15$). We show the \citet{stratevaColorSeparationGalaxy2001} $u-r=2.22$ separator as a dashed line above which galaxies are typically thought to be early-type, with late-type galaxies living below the separator. In the bottom right of each panel we show the fraction of objects that lie above the \citet{stratevaColorSeparationGalaxy2001} separator, while in the bottom left of each panel we show the fraction that lie below the separator. Most objects ($>$85\%) that do not satisfy our WISE SNR criteria are characterized by red colors and are found along the `red-sequence,' especially at higher redshifts. These objects are likely early-type galaxies with lower star-formation rates and diminished infrared emission, so called `red-and-dead' galaxies.\label{fig:sdsscolor}}
\end{figure*}

To investigate the effect of restricting our optical spectroscopic sample to having detected mid-IR photometry, we plot the SDSS $f_r/f_g$ versus $f_g/f_u$ of the SDSS Legacy MGS, our WISE Matched SNR sample, and our WISE SNR Rejected sample over two redshift regimes in Figure \ref{fig:sdsscolor}. We find that our WISE crossmatching and SNR cuts preferentially removes objects with high $f_r/f_g$ and $f_g/f_u$, i.e. red $g-r$ and $u-g$ colors, in both the low and high redshift regimes. In addition, we plot the \citet{stratevaColorSeparationGalaxy2001} $u-r=2.22$ separator which delineates late-type galaxies from early-type galaxies. We find that excluded objects predominantly lie along the `red-sequence' ($>$85\%). These galaxies are likely near the end of the galaxy evolutionary life cycle (e.g. `red-and-dead' galaxies) with lower star-formation rates and less infrared flux.

\subsection{SDSS Spectroscopy}\label{subsec:spectroscopy}

Our SDSS spectroscopy was observed with the original SDSS spectrographs which were fed with 3" diameter fibers and covered a wavelength range from 3800\AA$-$9200\AA\ with a resolution of $R\sim1800$ \citep{yorkSloanDigitalSky2000}. While we attempt to fit all of the 452,323 spectra in our WISE Matched SNR sample, this is comprised of many objects with duplicate observations.

\begin{figure*}[ht!]
    \includegraphics[width=\textwidth]{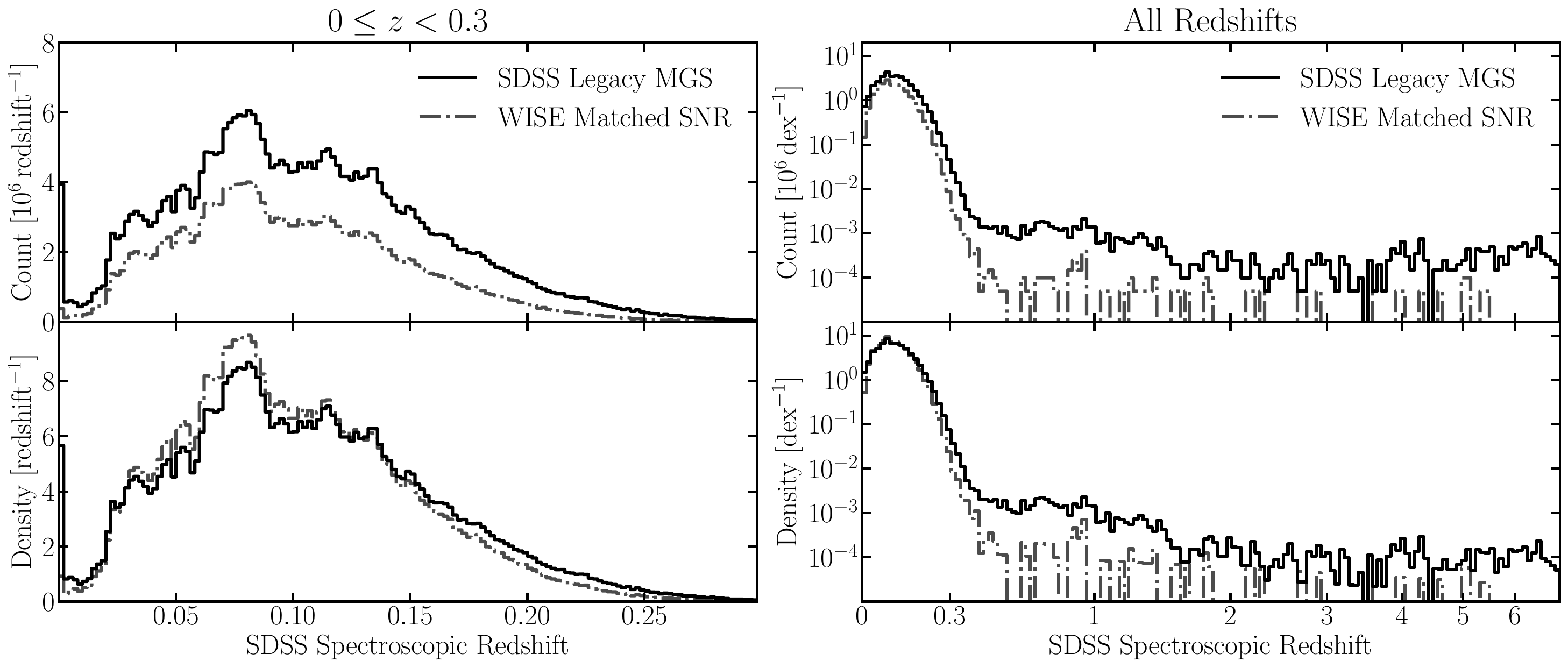}{}
    \caption{Redshift distributions of objects in the SDSS Legacy MGS and our WISE Matched SNR sample. In the left panels we show the distribution only for redshifts $0\leq z<0.3$ which make up the vast majority of the samples. The distribution across all redshifts is shown in the right panels. In the top panels we show the count histograms, while in the bottom panels we show the density histograms. The distribution for the SDSS Legacy MGS is shown with a solid line and the distribution for our WISE Matched SNR is shown with a dot-dashed line.\label{fig:redshift}}
\end{figure*}

In Figure \ref{fig:redshift}, we show the SDSS redshift distributions of the SDSS Legacy MGS and WISE Matched SNR sample across two redshift regimes. By matching to WISE and imposing our flux SNR criteria, we bias the WISE Matched SNR sample to brighter objects. In doing so, we bias the sample to a slightly lower redshift and especially reduce the number of high-redshift objects ($z >0.5$).

\subsection{SDSS Photometry}\label{subsec:sdss}

Our SDSS photometry is taken from the SDSS DR17. As of DR17, new photometric data and reduction pipelines have updated much of the photometry of the galaxies in the MGS. However, a small fraction of objects are left with no photometry as a consequence. As less than a $0.01\%$ of objects are left without full SDSS photometry, we use the improved DR17 photometric measurements for the MGS. We use the \texttt{dered} magnitude, which provides the model magnitude accounting for Milky Way extinction. The magnitudes are converted to a flux in Janskys and the corresponding flux error is generated from the model magnitude error (\texttt{modelMagErr}). If there is no photometry provided, i.e. the value of the magnitude is $-$9999, the band is not used in the SED fitting of the object. The vast majority of the WISE Matched SNR sample has coverage in all five bands (416,259 objects, 99.99\%). Two objects have measurements in only four bands, three objects have measurements in only three bands, five objects have measurements in only two bands, and 19 objects have no SDSS photometry in DR17.

\begin{figure}[ht!]
    \includegraphics[width=\columnwidth]{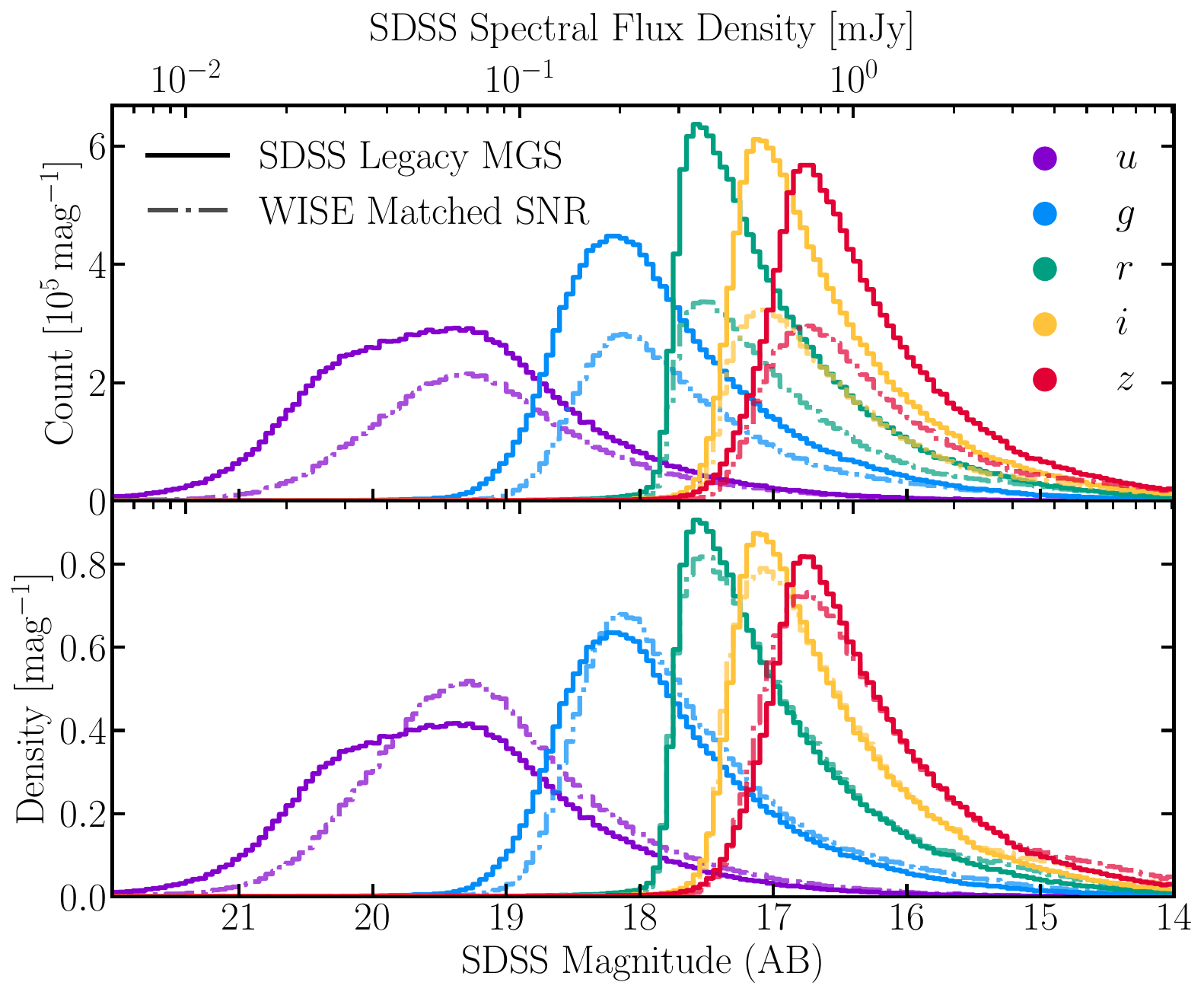}{}
    \caption{AB magnitude distributions of SDSS photometry in the SDSS Legacy MGS and our WISE Matched SNR sample. In the top panel we show the count histogram, while in the bottom panel we show the density histogram. The distributions for the SDSS Legacy MGS are shown with solid lines and the distributions for our WISE Matched SNR sample are shown with dot-dashed lines.\label{fig:sdssmag}}
\end{figure}

In Figure \ref{fig:sdssmag}, we show the distributions of the SDSS photometry for the SDSS Legacy MGS and WISE Matched SNR sample. Matching to WISE and applying our SNR cuts biases the WISE Matched SNR sample towards slightly brighter objects, as well as objects with a brighter $u$-band flux. As discussed earlier, our WISE selection preferentially excludes early-type galaxies, eliminating objects with lower star-formation rates, and therefore increasing the fraction of star-forming galaxies in the WISE Matched SNR sample, resulting in the brighter $u$-band flux distribution.

\begin{figure}[ht!]
    \includegraphics[width=\columnwidth]{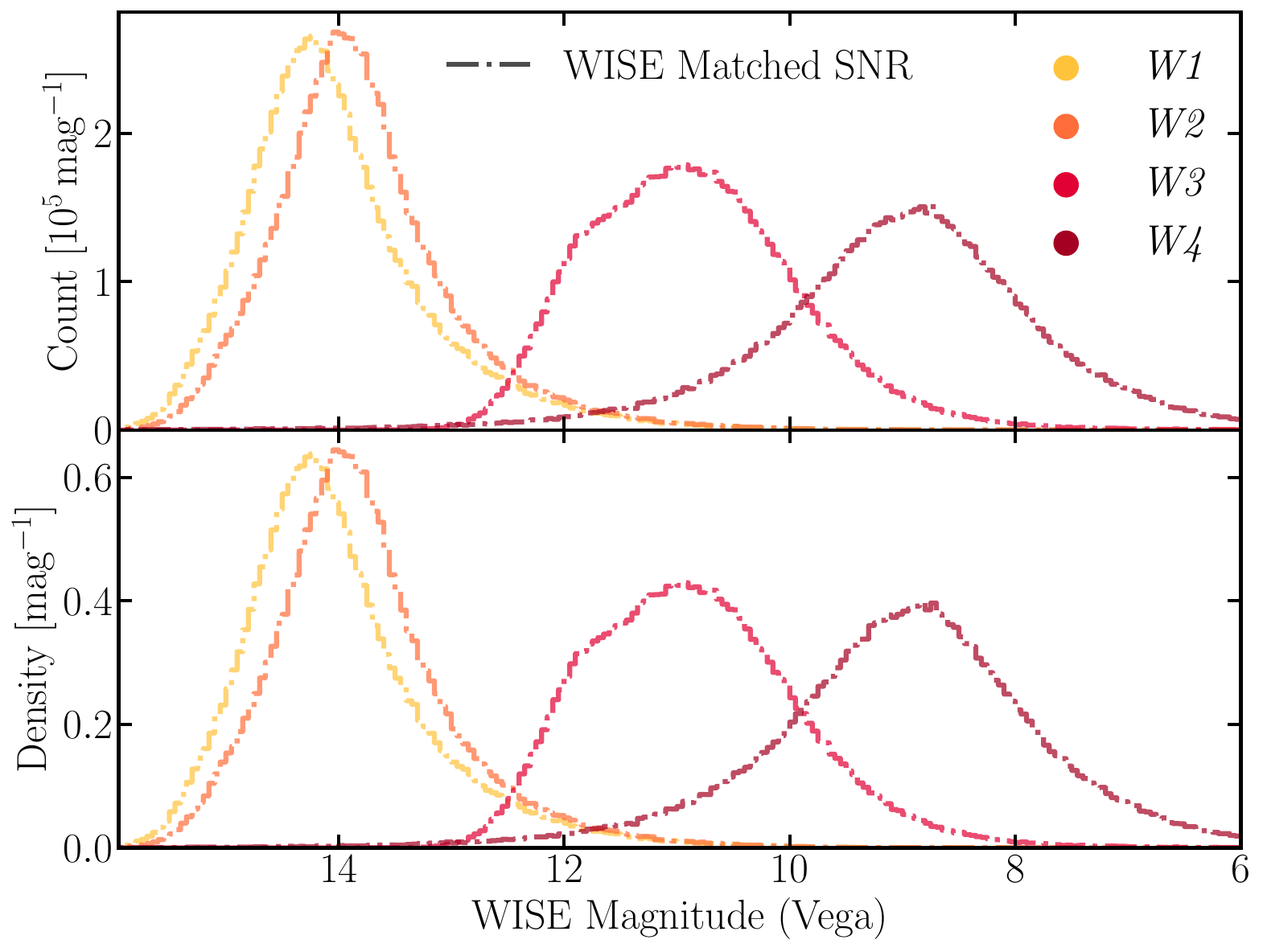}{}
    \caption{Vega magnitude distributions of the WISE photometry in our WISE Matched SNR sample. In the top panel we show the count histogram, while in the bottom panel we show the density histogram.\label{fig:wisemag}}
\end{figure}

\subsection{WISE Photometry}\label{subsec:wise}

The allWISE data catalog combines the cryogenic and post-cryogenic survey phases in the most comprehensive WISE release across all four survey bands and achieves a 95\% completeness above 0.04, 0.09, 0.8, and 7\,mJy in the \textit{W1}, \textit{W2}, \textit{W3}, and \textit{W4} bands respectively, making it the ideal choice for this work.

AllWISE provides a choice of photometric measurements to use, including profile fitting, aperture, and elliptical magnitudes, among others. To decide which photometry to use for each WISE band, we use the extended source flag (\texttt{ext\_flg}) following the suggestions in the Frequently Asked Questions section of the allWISE Explanatory Supplement \citep[aWES;][]{cutriExplanatorySupplementAllWISE2013}. The \texttt{ext\_flg} measurement describes whether the source is a point source or a resolved source and if the source is associated with a previously known source. Following information from the aWES, if the value of the flag is 0, 2, or 4, we use the profile fitting photometry (\texttt{mpro}). If the value of the flag is 1 or 3, we use the standard aperture magnitude (\texttt{mag}). Finally, if the flag has a value of 5, we use the elliptical magnitude (\texttt{gmag}). Each magnitude is converted into a flux in Janskys using the Vega zeropoints and correction factors corresponding to a flat spectrum in Table 1 of \citet{wrightWidefieldInfraredSurvey2010}. The associated magnitude error is then also converted into a flux error. These fluxes and errors are used to impose our 3$\sigma$ detection criterion on the WISE Matched SNR sample in \textit{W1}, \textit{W2}, and \textit{W3}. 

After applying the SNR cuts, if only an upper limit is provided in \textit{W4} for an object (i.e. a zero magnitude error), then the raw \textit{W4} source count from profile fitting is used (\texttt{w4flux}). This is the case for roughly half of the WISE Matched SNR sample (189,380 objects, 45.49\%). The count is converted into a flux in Janskys using the \textit{W4} instrumental zero point magnitude found in the aWES. The corresponding count error (\texttt{w4sigflux}) is converted to a flux error in the same way. However, if the value of the count error is zero for a given object, then no information is available in the \textit{W4} band and it is not used in the SED fitting of the object. Only six objects in the WISE Matched SNR sample have no information in the W4 band.

In Figure \ref{fig:wisemag}, we show the distributions of WISE photometry for our WISE Matched SNR sample.

\subsection{Summary \& Data Tables}\label{subsec:photometry}

We compile the names of the derived samples defined throughout Section \ref{sec:sample} along with the number of objects and corresponding spectra in Table \ref{tab:samples} for reference and easy comparison.

Our spectroscopic and photometric sample data are compiled and presented in a \texttt{FITS} table hosted on Zenodo:\dataset[doi:10.5281/zenodo.5848048]{https://doi.org/10.5281/zenodo.5848048}. The table includes information about the spectra provided by SDSS along with survey IDs, astrometric positions, and our final computed fluxes for each object used throughout the analysis. The description for each column of the data table can be found in Table \ref{tab:sample}.

\begin{deluxetable}{l|c|c}
    \tablecaption{Samples Presented in this Work\label{tab:samples}}
    \tablehead{
        \multicolumn{1}{c|}{Sample Name} &
        \multicolumn{1}{c|}{Objects} & 
        \multicolumn{1}{c}{Spectra} 
    }
    \startdata
    SDSS Legacy Main Galaxy Sample & 708,557 & 770,697 \\
    Spectroscopic sample & 702,993 & 764,229 \\
    WISE Crossmatched sample & 674,631 & 733,404 \\
    WISE SNR Rejected sample & 258,343 & 281,081\\
    WISE Matched SNR sample & 416,288 & 452,323 \\
    \enddata
\end{deluxetable}

\section{Spectroscopic Fitting}\label{sec:spec}

To fit the emission line spectra of the WISE Matched SNR sample, we developed the Galaxy/AGN Emission Line Anaylsis TOol (GELATO). The software is maintained on GitHub\footnote{\url{https://github.com/TheSkyentist/GELATO}} under a GPL-3.0 License is archived in Zenodo \citep{hvidingTheSkyentistGELATOGELATO2022}. For this work, we use GELATO version 2.5.1. While detailed fits to subsets of the SDSS MGS exist, we refit the WISE Matched SNR sample defined in Section \ref{sec:sample} and use our measurements of the emission line properties throughout this work to present conclusions derived in a consistent manner. To fit a spectrum, GELATO requires an initial estimate of the redshift. For this work, we use the redshift derived by the automated SDSS pipeline, $z_0$, as the initial estimate.

Due to limitations in GELATO, we cannot fit every possible spectrum. For example, GELATO is unable to fit the spectra of stars. In our WISE Matched SNR sample, 2,248 spectra ($0.5\%$) from 2,030 objects are identified as stars by the automated SDSS classification pipeline. This means we cannot obtain GELATO fits for 2,014 objects in the WISE Matched SNR sample that only have spectra identified as stars. However, 16 objects have spectra identified as galaxies by the pipeline in addition to those identified as stars. Upon visual inspection, 15 of these objects are indeed galaxies and therefore we use the results from their spectroscopic fitting in the remainder of the analysis. The remaining object, 1237673310969593977, does not have a galaxy spectrum and we do not obtain GELATO fits for this object's spectra.

In addition, due to the range of continuum models and the SDSS spectral coverage, GELATO can only fit spectra with a redshift of $z_0 \leq 1$. In our WISE Matched SNR sample, 34 spectra from 34 objects do not satisfy this criterion. Each of these spectra are identified as quasars by the automated SDSS classification pipeline. Two of these objects have additional spectra with an identified $z_0 \leq 1$ and, upon visual inspection, we find that the low-redshift solution appears to be correct. For the remaining 32 objects, we visually inspect each one and find that 23 are indeed $z_0>1$ broad line quasars. Due to the quality of their data, we are unable to classify the remaining nine objects which we list in Table \ref{tab:notqso}.

\begin{deluxetable}{cc}
    \tablecaption{Unclassified $z_0 > 1$ Objects\label{tab:notqso}}
    \tablehead{\multicolumn{2}{c}{SDSS objID}}
    \startdata
    1237660765919183163 & 1237656528920707373\\
    1237654600489697292 & 1237667293187080468\\
    1237652944767550576 & 1237655124468302048\\
    1237655369831284924 & 1237657628440461503\\
    1237667293187014904 & \\
    \enddata
\end{deluxetable}
 
After accounting for stars and high-redshift galaxies, we fit the remaining 450,041 spectra using GELATO. In this section we describe the operation of GELATO in detail (Section \ref{subsec:gelato}), present the results of our GELATO fits (Section \ref{subsec:results}), and compare our results to existing emission line flux measurements (Section \ref{subsec:compare}). Finally in Section \ref{subsec:class}, we describe the criterion for spectroscopic classification of objects in the WISE Matched SNR sample. 

\subsection{GELATO}\label{subsec:gelato}

GELATO is a flexible framework for fitting emission lines written in \texttt{python}. GELATO allows users to easily and intuitively specify the relationship between emission lines kinematics and to test for the inclusion of additional complex components, e.g. broad emission lines, blueshifted wings, or absorption features, that may be present in the spectrum. For example, narrow emission lines that come from the same physical source, such as the Balmer series, e.g. H$\alpha$ and H$\beta$ should have identical kinematics. Numerically, this corresponds to their line-of-sight velocities and velocity dispersions having the same value. Similarly, the [\ion{O}{3}]$\lambda\lambda$4959,5007\AA\ doublet would not only have identical kinematics, but their flux ratio is fixed due to their relative transition probabilities well established from atomic physics \citep{galavisAtomicDataIRON1997}. GELATO makes it straightforward to specify which lines are related physically and which physical parameters are assumed to be shared between different emission species.

To fit the spectra we make use of the High Performance Computing (HPC) clusters at the University of Arizona. With an average runtime of 6.5\,min per spectrum, we used $\sim$47,000 central processing unit (CPU) hours on the HPC Puma cluster to fit 450,041 spectra in the WISE SNR matched sample. While this can be taken as a benchmark for the performance of GELATO, we note that the runtime of GELATO can depend heavily on the input spectrum and the number of bootstrap iterations used to constrain errors on the fitted parameters as described in Section \ref{subsubsec:emiss}.

GELATO uses a Trust Region Reflective \citep[TRF]{branchSubspaceInteriorConjugate1999} algorithm with a least-squares loss function throughout the fitting process. The algorithm accepts an initial estimate of the best-fit model parameters and iterates towards the best-fit solution. The TRF algorithm is ideal for use with GELATO since it is able to accept bounds on fitted parameters and is designed for problems with many variables.

\subsubsection{Continuum Fitting}\label{subsubsec:cont}

Modeling the stellar continuum is essential to retrieving accurate emission line fluxes. The optical continuum in galaxies is dominated by stars and can be modeled by synthesizing a combination of stellar populations at different ages and metallicities. GELATO models the continuum as a combination of Simple Stellar Populations (SSPs) from the Extended MILES stellar library \citep[E-MILES]{vazdekisUVextendedEMILESStellar2016}. As the goal of GELATO is to measure the emission line parameters in the WISE Matched SNR sample, SSP models are an ideal choice for reproducing the large range of continua present in SDSS spectroscopy while not requiring expensive stellar population synthesis. We use SSP models that assume a Chabrier IMF (slope $=1.3$) and isochrones of \citet[][commonly referred to as Padova+00]{girardiEvolutionaryTracksIsochrones2000} with solar alpha abundance. We select a subset of the SSP models to span a range of representative metallicities and ages ([M/H] = [$-$1.31, $-$0.40, 0.00] and Age = [00.0631, 00.2512, 01.0000, 04.4668, 12.5893] (Gyr)) with nominal resolutions of 5\AA. As the E-MILES models have a minimum wavelength of 1680\AA, there is a maximum redshift where the template will no longer overlap the input spectrum. For SDSS MGS spectroscopy this corresponds to a redshift of $z\simeq1.3$. 

To separate the continuum flux from the emission lines, GELATO begins by reading the provided line list, described in Section \ref{subsubsec:emiss}, and masks the spectrum in a user specified region around each line. To fully mask emission lines, including potential broad components, we mask a region of 10,000\,km\,s$^{-1}$ around each line. The E-MILES templates are redshifted to match up with the spectrum, interpolated to the spectrum wavelength, multiplied by normalization coefficients, and are summed together. The TRF algorithm is used to fit the continuum model to the masked continuum and retrieve the fitted continuum redshift and normalization coefficients. The coefficients of the E-MILES templates are constrained to be positive while the continuum redshift is allowed to vary within a range around the input redshift ($z_0 \pm 150$\,km\,s$^{-1}$).

In an unobscured AGN, temperatures in the accretion disk are sufficient to produce optical emission, commonly modeled as a power law of the form: $f \propto\lambda^{-\alpha}$. To account for this possibility, GELATO adds an additional power-law component following the initial fit. Adding a new model component will always result in a ‘better fit,’ i.e. a  lower $\chi^2$, due to the added degrees of freedom. To determine if the inclusion of the power law is statistically significant, we perform an F-Test to determine whether the relative increase in the degrees of freedom is justified by the relative decrease in the $\chi^2$. By comparing to an F-distribution, a likelihood that the extended model is supported by the data is generated. We again use a 3$\sigma$ confidence threshold, corresponding to a likelihood of $\sim$99.87\% on the F-Test. If the continuum model with a power law passes the F-Test, it is added to the model.

The continuum redshift of the model is frozen for the remainder of the fitting. By fixing the continuum redshift, the E-MILES templates do not have to be re-interpolated at every iteration. This reduces the time it takes to fit by not only reducing the number of operations at each iteration, but also makes the Jacobian of the final model analytic and directly calculable. Understanding the emission lines does not require a detailed continuum redshift measure; fixing it after fitting the masked continuum ensures a stable and accurate result. After fitting the final continuum model to the masked spectrum, if the TRF algorithm determines that any of E-MILES coefficient parameters are at their lower bound of zero, the corresponding E-MILES models are removed from the continuum model.

\begin{centering}
\begin{table*}
\centering
\caption{GELATO Emission Line Dictionary\label{tab:GELATO}}
\hspace{-2.35cm}
\noindent
\begin{tabular}{c|cc|ccccc}
\hline\hline\hline
\multicolumn{5}{c|}{\textbf{Species (Tie \boldmath$z$ \& $\sigma$)}} & \multicolumn{3}{c}{\textbf{Groups}}\\
\textbf{Name} & \multicolumn{1}{|c}{\textbf{Line [\AA]}} & \multicolumn{1}{c|}{\textbf{Ratio}} & \textbf{+Comp} & \multicolumn{1}{c|}{\textbf{$\to$Group}} & \textbf{Name} & \textbf{Tie \boldmath$z$?} & \textbf{Tie \boldmath$\sigma$?} \\\hline\hline
\multirow{2}{*}{[SII]}&6716.44 & -&\multirow{2}{*}{{\footnotesize }} & \multirow{2}{*}{{\footnotesize }}&\multirow{14}{*}{Narrow} & \multirow{14}{*}{True $\mid$ \textbf{False}} & \multirow{14}{*}{True $\mid$ \textbf{False}}\\
\cline{2-3}
&6730.82 & -\\
\cline{0-4}
\multirow{2}{*}{[NII]}&6583.45 & 1&\multirow{2}{*}{{\footnotesize }} & \multirow{2}{*}{{\footnotesize }}&\\
\cline{2-3}
&6548.05 & 0.34\\
\cline{0-4}
\multirow{2}{*}{[OI]}&6300.3 & 1&\multirow{2}{*}{{\footnotesize }} & \multirow{2}{*}{{\footnotesize }}&\\
\cline{2-3}
&6363.78 & 0.333\\
\cline{0-4}
\multirow{1}{*}{HeI}&5875.62 & -&\multirow{1}{*}{{\footnotesize }} & \multirow{1}{*}{{\footnotesize }}&\\
\cline{0-4}
\multirow{2}{*}{[OIII]}&5006.84 & 1&\multirow{2}{*}{{\footnotesize Outflow}} & \multirow{2}{*}{{\footnotesize Outflow}}&\\
\cline{2-3}
&4958.91 & 0.35\\
\cline{0-4}
\multirow{1}{*}{[OIII]}&4363.21 & -&\multirow{1}{*}{{\footnotesize }} & \multirow{1}{*}{{\footnotesize }}&\\
\cline{0-4}
\multirow{1}{*}{[NeIII]}&3868.76 & -&\multirow{1}{*}{{\footnotesize }} & \multirow{1}{*}{{\footnotesize }}&\\
\cline{0-4}
\multirow{2}{*}{[OII]}&3726.03 & -&\multirow{2}{*}{{\footnotesize }} & \multirow{2}{*}{{\footnotesize }}&\\
\cline{2-3}
&3728.82 & -\\
\cline{0-4}
\multirow{1}{*}{[NeV]}&3425.88 & -&\multirow{1}{*}{{\footnotesize }} & \multirow{1}{*}{{\footnotesize }}&\\
\cline{0-7}
\multirow{1}{*}{HI}&6562.79 & -&\multirow{1}{*}{{\footnotesize Broad}} & \multirow{1}{*}{{\footnotesize Broad}}&\multirow{3}{*}{Balmer} & \multirow{3}{*}{\textbf{True} $\mid$ False} & \multirow{3}{*}{\textbf{True} $\mid$ False}\\
\cline{0-4}
\multirow{1}{*}{HI}&4861.28 & -&\multirow{1}{*}{{\footnotesize Broad}} & \multirow{1}{*}{{\footnotesize Broad}}&\\
\cline{0-4}
\multirow{1}{*}{HI}&4340.47 & -&\multirow{1}{*}{{\footnotesize Broad}} & \multirow{1}{*}{{\footnotesize Broad}}&\\
\cline{0-7}
    &  &  &  &  & \multirow{1}{*}{Outflow} & \multirow{1}{*}{True $\mid$ \textbf{False}} & \multirow{1}{*}{True $\mid$ \textbf{False}}\\
\cline{0-7}
    &  &  &  &  & \multirow{1}{*}{Broad} & \multirow{1}{*}{\textbf{True} $\mid$ False} & \multirow{1}{*}{True $\mid$ \textbf{False}}\\
\cline{0-7}
\hline\hline\hline
\end{tabular}
\tablecomments{This table demonstrates the Emission Line Dictionary inputted into GELATO for the fitting in this work. The table also serves as an example of the hierarchical structure of GELATO described in Section \ref{subsubsec:emiss}. Emission lines, defined by a wavelength, are grouped into \textit{Species}. Lines in \textit{Species} must share kinematics. \textit{Species} are grouped into \textit{Groups}. \textit{Groups} can optionally share kinematics. Additional components can be added to \textit{Species} along with corresponding target \textit{Groups}.}
\end{table*}
\end{centering}

\subsubsection{Emission Line Fitting}\label{subsubsec:emiss}

GELATO uses a hierarchical approach to represent the physical relationship between emission line species and components. GELATO first associates individual emission lines, characterized by a central wavelength, with a \textit{Species}. For example, all Balmer emission lines would be characterized by a single \textit{Species}, e.g. \ion{H}{1}. Since all emission lines within a \textit{Species} come from the same physical source, their kinematics are tied together, e.g. their redshifts and velocity dispersions are required to be the same value, respectively. In addition, a flux ratio can be specified for emission lines whose relative intensities are fixed. The various \textit{Species} are then distributed into \textit{Groups}. \textit{Groups} embody assumptions made between their various component \textit{Species}. The user can decide whether \textit{Species} in a \textit{Group} are required to have their velocity dispersion or redshifts fixed to the same value, respectively. In summary, \textit{Groups} contain \textit{Species}, which can optionally share kinematic properties, while \textit{Species} contain emission lines that must share kinematic properties. 

For each \textit{Species}, the user can also specify which additional components should be tested for their inclusion. Any combination of the supported additional components can be specified for all \textit{Species}. For each additional component specified, a \textit{Group} must be specified that the additional component will be placed into. This allows for the additional component to have a different relationship with other \textit{Species} than the parent \textit{Species}. As a consequence, it allows the user to specify whether or not additional components will have a velocity shift with respect to their narrow counterparts.

In Table \ref{tab:GELATO} we present the GELATO configuration used in this work. It displays the entire suite of emission lines fit in this work along with the additional components which will be tested. For the [\ion{O}{3}] emission complex, an additional outflow component will be tested which, if accepted, is not forced to share the redshift or dispersion of the narrow components. For the Balmer lines, a broad additional component will be tested which, if accepted, is not forced to share the redshift or dispersion of the narrow components. We note, however, that the broad Balmer components are also not forced to share a dispersion value, as systematic offsets between the broad Balmer lines dispersions have been observed in powerful quasars \citep[e.g.][]{greeneEstimatingBlackHole2005}.

In GELATO, each emission line is treated as a Gaussian parametrized by its redshift, velocity dispersion (the standard deviation of the Gaussian), and flux. GELATO determines which input emission lines are covered by the spectrum by using the initial estimate of the redshift from SDSS. The spectrum must contain data within a width of 300\,km\,s$^{-1}$ centered around a given emission line central wavelength for it to be included. The initial model is then constructed with each emission line as a single Gaussian and the continuum model. The flux has symmetric bounds based on the maximum flux possible based on the estimated height of the emission line to prevent non-physical values.

Line kinematics in AGNs can differ from a single narrow Gaussian function. High velocity dispersion gas near the nucleus or bulk motion of ionized gas due to the influence of the AGN can cause more complex line shapes. By adding additional Gaussian components, the kinematics of these emission complexes can be modeled. GELATO is designed to test if any additional components are necessary to fit the complex emission lines. To determine if an additional component is statistically significant, GELATO fits a spectrum with each potential additional component in turn. The new component must satisfy an F-Test at the 3$\sigma$ level, corresponding to a likelihood of $\sim$99.87\%, to be accepted.

A model is then computed for all possible combinations of accepted additional components. Each of these is fit to the spectrum and the Akaike Information Criterion \citep[AIC]{akaikeNewLookStatistical1974} is then calculated for each one. The model with the lowest AIC is selected as the final GELATO model for the spectrum. However, if the TRF algorithm determines that any parameters of the new components in the combination model are at their bounds, i.e. at the edge of the ranges described in Section \ref{subsubsec:values}, the combination model will not be selected as the final model.

The final model is then fit to the spectrum using the TRF algorithm. To determine the errors on the fitted parameters, GELATO re-generates the spectrum flux within the uncertainties 500 times and each synthetic spectrum is refit. The standard deviation across all 500 `bootstraps' in each parameter is taken as its uncertainty and the median of each parameter is taken as its final value.

\subsubsection{Emission Line Parameter Value Ranges}\label{subsubsec:values}

We base the range of parameter values of the narrow emission lines on the AGN Line Profile And Kinematics Archive \citep[ALPAKA]{mullaneyNarrowlineRegionGas2013} which fit multiple Gaussian components to narrow emission lines. The redshift is constrained within a range consistent with over 98\% of the velocity shifts of narrow lines from ALPAKA: $z_0 \pm 300$\,km\,s$^{-1}$. The dispersion has a lower bound of 60\,km\,s$^{-1}$, consistent with maximum resolution of the original SDSS spectrograph \citep{yorkSloanDigitalSky2000} and an upper bound of 500\,km\,s$^{-1}$ corresponding to the delineation between narrow lines and broad lines at a FWHM of 1200\,km\,s$^{-1}$ used in \citet{haoActiveGalacticNuclei2005}. The dispersion has a starting value of 130\,km\,s$^{-1}$ based on the median dispersion of narrow lines from ALPAKA.

GELATO supports two additional components: a blueshifted component represented outflowing gas and a broad line component. As these additional components represent sources of line emission, we constrain the flux to be positive with an upper bound based on the estimated height of the emission line.

We base the range of parameter values of the outflow additional component on ALPAKA. The redshift of the outflow component has a lower bound of $z_0-$750\,km\,s$^{-1}$, an upper bound of $z_0+$150\,km\,s$^{-1}$, and a starting value of $z_0-$150\,km\,s$^{-1}$ based on the 1, 99, and 50 percentile values of second [\ion{O}{3}] component invoked in the ALPAKA fitting. The dispersion of the outflow component has a lower bound of 100\,km\,s$^{-1}$, an upper bound of 750\,km\,s$^{-1}$, and a starting value of 300\,km\,s$^{-1}$ based on the 1, 99, and 50 percentile values of second [\ion{O}{3}] component invoked in the ALPAKA fitting.

We base the range of parameter values of the broad line additional component on the \citet[S11]{shenCatalogQuasarProperties2011} study of SDSS optical quasars. The redshift of the broad component is constrained within a range consistent with over 90\% of the velocity shifts of the broad H$\alpha$ and H$\beta$ lines from the S11 study: $z_0 \pm 600$\,km\,s$^{-1}$. The broad line dispersion has a lower bound of 500\,km\,s$^{-1}$ corresponding to the delineation between narrow lines and broad lines at a FWHM of 1200\,km\,s$^{-1}$ used in \citet{haoActiveGalacticNuclei2005}. The dispersion of the broad line has an upper bound of 6500\,km\,s$^{-1}$ and a starting value of 1600\,km\,s$^{-1}$ based on the 99 and 50 percentile values of the broad H$\alpha$ and H$\beta$ lines measured in the S11 fitting.

\begin{figure*}[ht!]
    \centering
    \includegraphics[height=0.45\textheight]{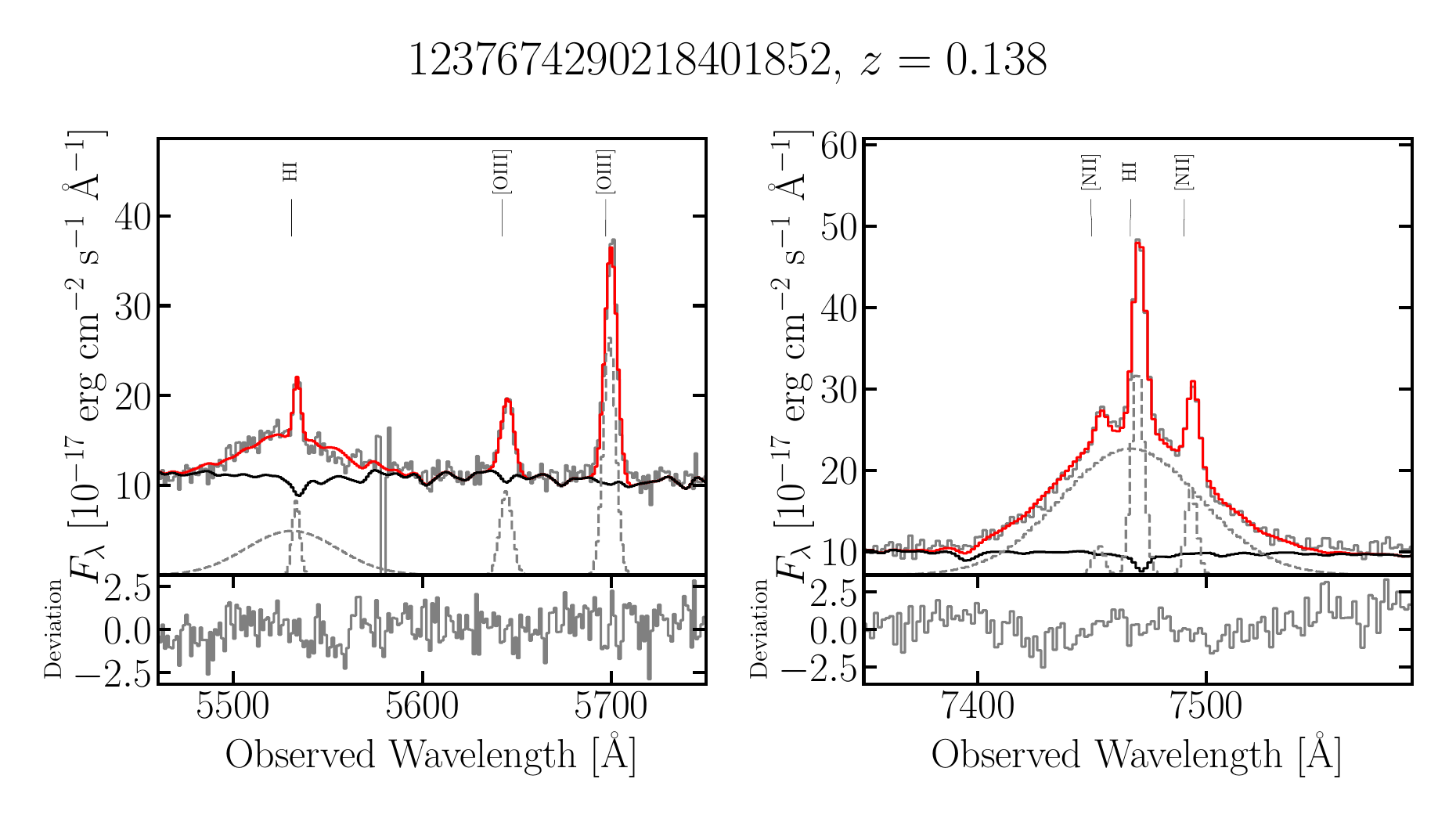}{}
    \includegraphics[height=0.45\textheight]{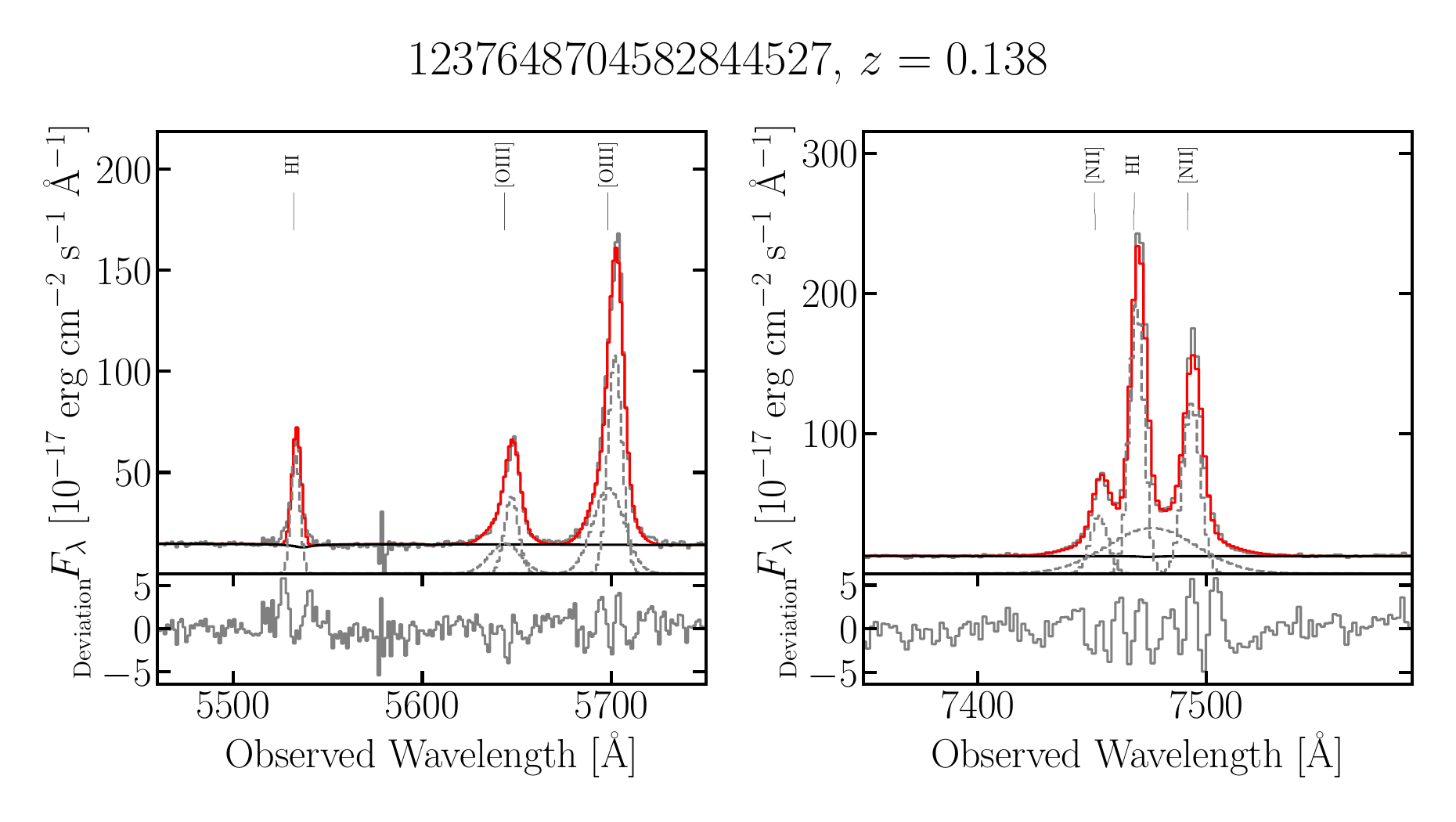}{}
    \caption{GELATO spectroscopic fits for SDSS objIDs 1237674290218401852 (top figure) and 1237648704582844527 (bottom figure). In both instances, we show the [\ion{O}{3}]+H$\beta$ and [\ion{N}{2}]+H$\alpha$ complexes for the objects. The spectrum is plotted as a gray line, the components are plotted as a gray dashed line, and the overall fit is shown in red. Below each fit, the deviation (residual divided by error) is shown in gray. GELATO is able to model the additional components present in the emission lines. Specifically, GELATO can model the broad components of the Balmer lines and the blueshifted components in [\ion{O}{3}].\label{fig:specsample}}
\end{figure*}

\subsection{GELATO Results}\label{subsec:results}

In this section we enumerate the results of our GELATO fitting and any additional quality cuts we place on the data to make them suitable for further analysis. While the TRF algorithm is ideal for a problem of this nature, it has one noticeable weakness. Due to the fact that the TRF algorithm only generates iterations that are strictly within the given constraints, it is not able to robustly determine when parameters are at their bounds, i.e. at the edge of the ranges described in Section \ref{subsubsec:values}. Since GELATO uses the TRF algorithm to reject additional components during emission line fitting, it may not always be able to reject all spurious additional components. Therefore, it is required that we impose additional quality cuts on the final data to ensure a robust detection of additional components.

To classify an additional component as a detection, it must pass the F-Test in GELATO at the 3$\sigma$ level. To ensure it is not a spurious additional component, it is not classified as a detection if its dispersion or velocity parameter comes within 0.01\,km\,s$^{-1}$ of its upper or lower limits as defined in Section \ref{subsubsec:values}. Finally, to ensure we are selecting on the presence of a broad line, and not simply an additional component which may have been required to fit the wings of nearby lines or stellar continuum features, we require that the flux in the broad additional component be at least 50\% of the flux in the narrow component for classification, similar to \citet{dipompeoIIIProfilesInfraredselected2018}. Since these additional quality cuts are imposed following the GELATO fitting, not classifying an additional component as a detection does not change the GELATO results or the retrieved parameters on the fit.

We demonstrate the ability of GELATO to fit emission line complexes, especially those with broad lines or outflowing components, by looking at two example objects, SDSS objIDs 1237674290218401852 and 1237648704582844527. We present the GELATO fits for the [\ion{O}{3}]+H$\beta$ and [\ion{N}{2}]+H$\alpha$ complexes of the objects in Figure \ref{fig:specsample}. Both show evidence for the presence of broad H$\alpha$ line, i.e. GELATO used an F-Test to determine that the inclusion of a broad line resulted in a statistically significant better fit. However, GELATO includes a broad H$\beta$ component for 1237674290218401852 and an outflowing [\ion{O}{3}] component for 1237648704582844527.

For the remainder of the analysis, we use the BPT diagnostic diagram to analyze the WISE Matched SNR sample. To robustly place galaxies onto this diagram, we require a flux SNR $>3$ in each of the BPT emission lines, i.e. H$\alpha$, H$\beta$, [\ion{N}{2}]$\lambda$6583.45\AA, and [\ion{O}{3}]$\lambda$5006.84\AA. This is equivalent to the quality cut placed on the detection of additional components with GELATO, ensuring a consistent detection quality across both broad and narrow components.

With the added quality cuts on the additional components, we identify 14,524 (3.5\%) galaxies with a broad line in H$\alpha$, 3,424 (0.8\%) galaxies with a broad line in H$\beta$, and 18,686 (4.5\%) galaxies with an outflowing [\ion{O}{3}] component. In addition, we identify 226,439 (54.4\%) galaxies with a flux SNR $>3$ in each of the BPT emission lines.

The spectroscopic results table is available on Zenodo:\dataset[doi:10.5281/zenodo.5848048]{https://doi.org/10.5281/zenodo.5848048} and the columns are described in the GELATO documentation\footnote{\url{https://github.com/TheSkyentist/GELATO\#readme}}. For our galaxies with multiple spectroscopic observations we only use a single spectroscopic result to not bias our results. For galaxies with two or more spectroscopic observations, we select the spectroscopic result with the $\chi^2_\nu$ closest to one. We do this rather than averaging to avoid any true physical difference between the spectra in the case of a differing fiber placement or a change in the object over time between the two observations.

\begin{figure*}[ht!]
    \centering
    \includegraphics[width=\textwidth]{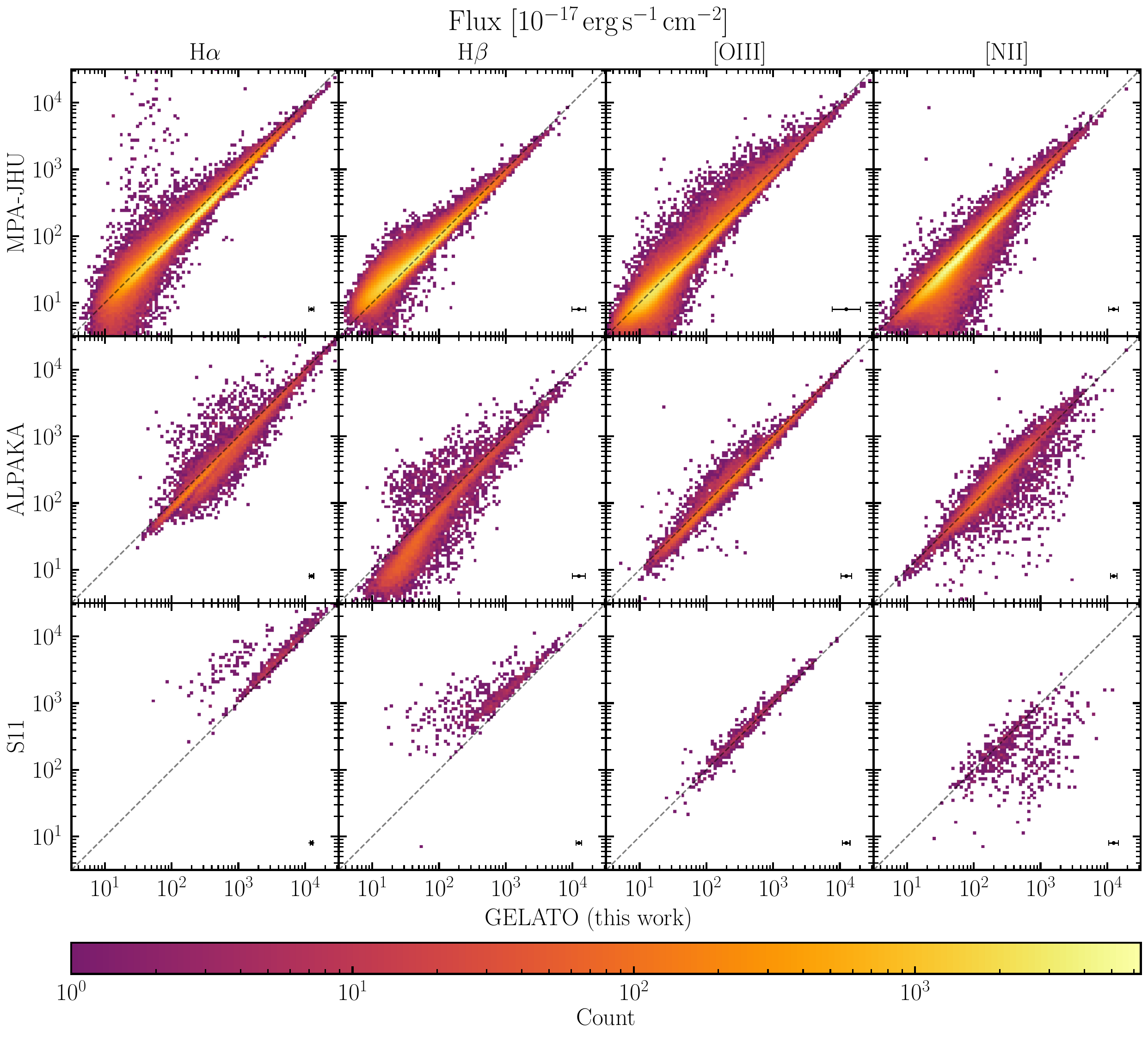}{}
    \caption{BPT emission line fluxes compared between GELATO and the MPA-JHU catalog, ALPAKA, and the S11 study. For each line we restrict to objects that have at least one component detected at the 3$\sigma$ level with GELATO. For the MPA-JHU catalog we only compare the narrow components while for the remaining studies we compute the total GELATO flux in each line as a sum of all detected components. For each line, the GELATO fluxes are are compared to the total flux from all line components for each study. In each panel we show error bars corresponding to three times the median GELATO uncertainty in the bottom right corner.\label{fig:speccomp}}
\end{figure*}

\begin{figure}[ht!]
    \includegraphics[width=\columnwidth]{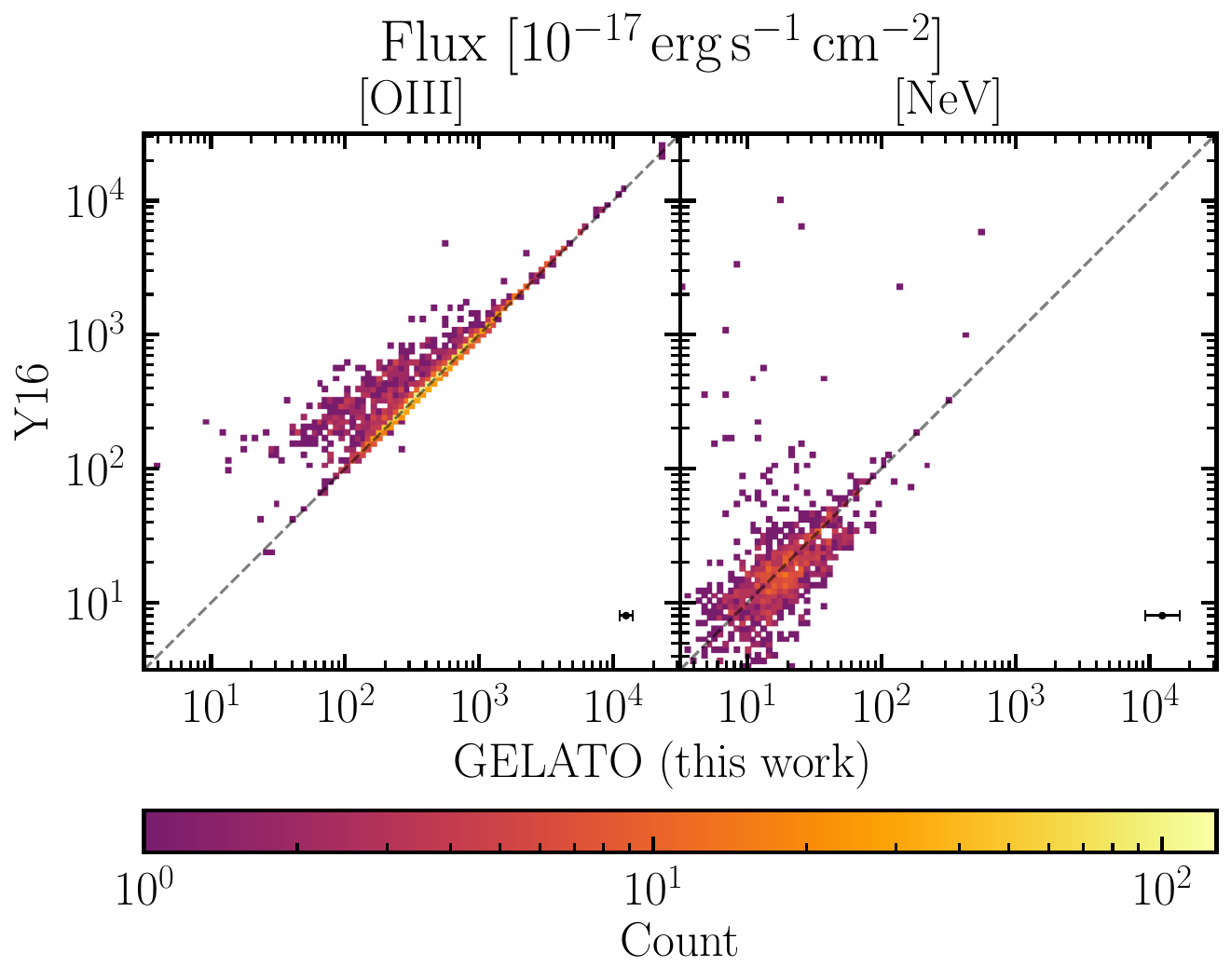}{}
        \caption{BPT [\ion{O}{3}] and [\ion{Ne}{5}]$\lambda3426.85$\AA\ emission line fluxes compared between GELATO and the Y16 study. For each line we restrict to objects that have at least one component detected at the 3$\sigma$ level with GELATO. We compute the total GELATO flux in each line as a sum of all detected components. For each line, the GELATO fluxes are compared to the total flux from all Y16 line components. In each panel we show error bars corresponding to three times the median GELATO uncertainty in the bottom right corner.\label{fig:Y16}}
\end{figure}

\subsection{Comparison to other Spectroscopic Fitting}\label{subsec:compare}

To evaluate the success of GELATO to retrieve emission line fluxes for SDSS spectroscopy, we compare to four catalogs of SDSS spectroscopic fits in the literature. We compare to the MPA-JHU catalog \citep[methods described in][]{brinchmannPhysicalPropertiesStarforming2004,kauffmannStellarMassesStar2003,tremontiOriginMassMetallicityRelation2004}, the ALPAKA catalog, the S11 quasar catalog, and the \citet[Y16]{yuanSpectroscopicIdentificationType2016} Type II quasar catalog. The MPA-JHU catalog, the ALPAKA catalog, and the S11 study all performed analysis on SDSS DR7 \citep{abazajianSeventhDataRelease2009}, while the Y16 study performed analysis on SDSS DR12 \citet{alamEleventhTwelfthData2015}. As the Y16 catalog was selected to target galaxies outside the Legacy MGS, it has no matches with our WISE Matched SNR sample, we refit the entire sample of 2,920 DR12 spectra with GELATO using the HPC El Gato cluster.

We only compare lines which are detected at the 3$\sigma$ level in the GELATO results. As the MPA-JHU study only used single Gaussian to measure emission line fluxes, we only compare the narrow components of the GELATO fits if the GELATO lines are detected with a flux SNR $>3$. For the remaining studies, we compare objects in a given line in which at least one GELATO component is detected at the 3$\sigma$ level corresponding to a flux SNR $>3$ for the narrow component, or the detection criteria outlined in Section \ref{subsec:results} for additional components. We compute the total GELATO flux in each line as a sum of all detected components and compare them to the total flux in each line summed across the provided components for each study.

In Figure \ref{fig:speccomp} we compare all four BPT line fluxes for the MPA-JHU catalog, ALPAKA, and the S11 study. As Y16 only provides estimates of the BPT [\ion{O}{3}] and [\ion{Ne}{5}]$\lambda3426.85$\AA\ lines, we present the comparisons separately in Figure \ref{fig:Y16}. In Table \ref{tab:compare} we compile the results from our spectroscopic comparisons by computing the median difference between the GELATO line fluxes and each of our comparison studies. We find that GELATO is usually within 0.1\,dex of the comparison studies with a few exceptions. In general, agreement is worse in the low flux limit where line strength and SNR is at its weakest. In the S11 study a clear bimodality can be observed in H$\alpha$ and H$\beta$ where GELATO cannot identify the broad line at the 3$\sigma$ level, resulting in enhanced [\ion{N}{2}] emission due to the proximity of this line to H$\alpha$. In addition, we find disagreements in measurements of the H$\beta$, which may be driven by differences in the continuum fitting between as the H$\beta$ line suffers from the most relative stellar continuum contamination.

In summary, we find that the GELATO results are broadly consistent with the previous existing emission line measurements. We see disagreement mostly in the case of Balmer line fitting. However, this can be a result of different studies using different number of components for fitting. Even with this, our results are usually consistent to within 0.1\,dex. GELATO thus provides an accurate measurement of emission line fluxes for SDSS spectroscopy.

\begin{deluxetable*}{c|c|c|c|c|c|c}
    \tablecaption{Spectroscopic Comparisons to GELATO\label{tab:compare}}
    \tablehead{
        \multirow{2}{*}{Study} & \multirow{2}{*}{Matched} & \multicolumn{5}{c}{Median Line Flux Difference [dex]}  \\
        & & \multicolumn{1}{c}{H$\alpha$} & \multicolumn{1}{c}{H$\beta$} & \multicolumn{1}{c}{[\ion{O}{3}]} & \multicolumn{1}{c}{[\ion{N}{2}]} & \multicolumn{1}{c}{[\ion{Ne}{5}]}
    }
    \startdata
    MPA-JHU & \hfill433,718 & $+$0.07 & $-$0.01 & $+$0.07 & $+$0.11 & --- \\
    ALPAKA & \hfill15,255 & $+$0.09 & $+$0.22 & $+$0.05 & $+$0.03 & --- \\
    S11 & \hfill621 & $-$0.06 & $-$0.19 & $-$0.01 & $+$0.22 & --- \\
    Y16 & \hfill2,920 & --- & --- & $\sim$0.00 & --- & $+$0.05 \\    
    \enddata
\end{deluxetable*}

\subsection{Classification}\label{subsec:class}

Using the GELATO results from fitting SDSS spectroscopy, we can place each object into an optical spectroscopic class. We first define a class of optical Type I AGN (T1), characterized by well-detected broad Balmer emission lines. A galaxy is placed into the T1 class if it has a broad H$\alpha$ or H$\beta$ line as defined in Section \ref{subsec:results}. In addition, we include the 23 $z_0>1$ quasars identified in Section \ref{subsec:spectroscopy} in the T1 class.

We can then classify using the BPT diagram. For a galaxy to receive BPT spectroscopic classification it must have a flux SNR $>3$ in all of the requisite lines for the BPT diagram. Those galaxies which lie above the \citet{kewleyTheoreticalModelingStarburst2001} demarcation are classified as optical Type II AGNs (T2), those that lie above the \citet{kauffmannHostGalaxiesActive2003} demarcation but below the \citet{kewleyTheoreticalModelingStarburst2001} line are classified as composite galaxies (Comp), and the remaining are classified as star-forming galaxies (SF).

Finally, the 2,014 objects in Section \ref{subsec:spectroscopy} that only have spectra identified as stars are placed in the Stars category. As few stars are identified in the WISE Matched SNR sample, with less than ten being selected by any WISE mid-IR criterion, they have little to no effect on understanding the selection of AGN from mid-IR photometry matched to optical data.

Galaxies which cannot be placed into any of the four categories defined above are denoted as `Unclassified' objects. An object may be Unclassified for a variety of reasons. For example, the galaxy may not host an AGN or star formation and therefore lack spectral lines or the activity may be sufficiently faint that the lines cannot be robustly detected. Even in the case where lines can be robustly detected, it may be that the spectrum is at a sufficiently high redshift that we no longer have coverage of all four BPT lines. Therefore, for these objects, we are unable to classify objects from spectroscopy alone. For the purposes of this work, these objects will not count as AGNs, and therefore any statistics reporting the numbers of selected AGNs will be lower limits.

A total of 232,288 galaxies in our WISE Matched SNR sample (55.8\%) can be placed into one of the five classifications. In Table \ref{tab:class} we summarize the number of objects that fall into each class.

\begin{deluxetable}{l|rr}
    \tablecaption{Classifications Presented in this Work\label{tab:class}}
    \tablehead{
        \multicolumn{1}{c|}{Classification} &
        \multicolumn{2}{c}{Galaxies} 
    }
    \startdata
    Optical Type I AGN & 14,658 & (3.5\%) \\
    Optical Type II AGN & 21,055 & (5.1\%) \\
    Composite Galaxies (Comp) & 49,969 & (12.0\%) \\
    Star-forming Galaxies (SF) & 144,592 & (34.7\%) \\
    Stars & 2,014 & (0.5\%) \\
    Unclassified & 184,000 & (44.2\%)
    \enddata
\end{deluxetable}

\begin{figure*}[ht!]
    \centering
    \includegraphics[height=0.45\textheight]{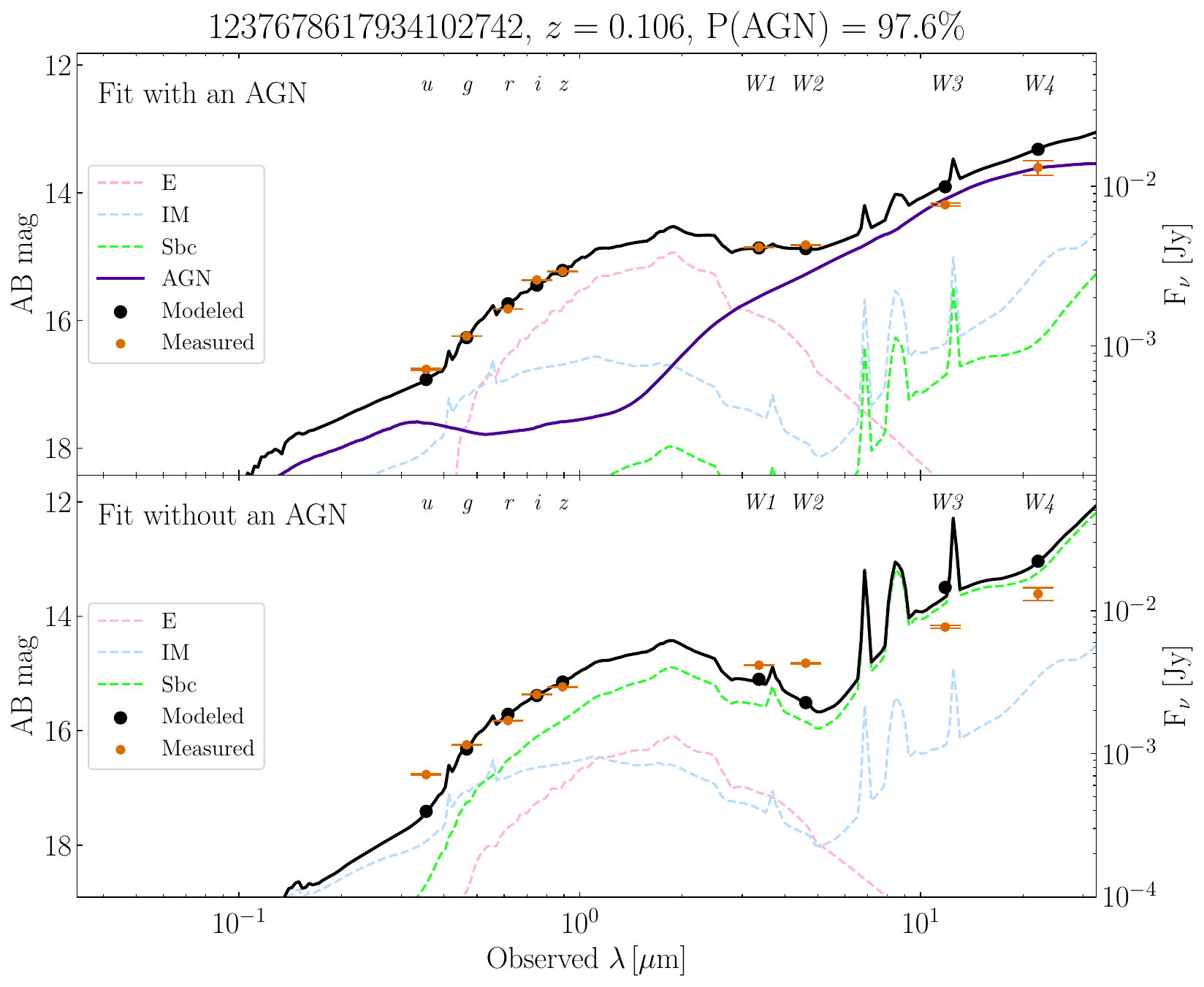}{}
    \includegraphics[height=0.45\textheight]{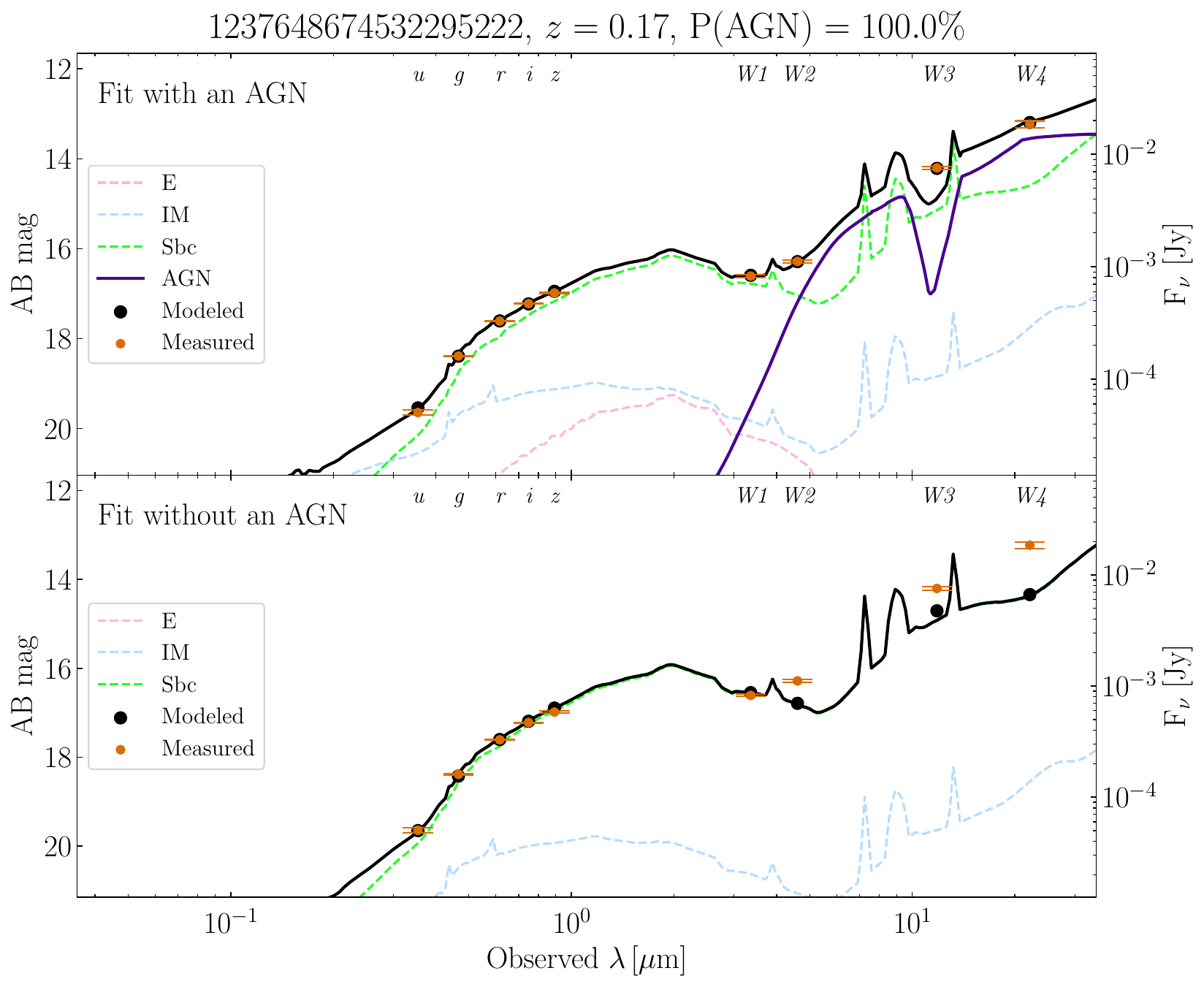}{}
    \caption{SED modeling for SDSS objIDs 1237678617934102742 (top figure) and 1237648674532295222 (bottom figure). In both instances, we show the SED fit with/without and AGN component with an extinction curve in the top/bottom panel. Both objects show strong evidence for requiring the AGN component in their modeling. In both objects, the \textit{W1$-$W2} color cannot be explained by the galaxy templates alone. While 1237678617934102742 shows evidence for an unobscured AGN, owing in part to an under-prediction of the $u$ band flux by the galaxy templates, 1237648674532295222 is best fit by an obscured quasar, where there is little AGN emission visible in the optical.\label{fig:photexample}}
\end{figure*}

\section{SED Analysis}\label{sec:sed}

SED analysis is a useful tool for determining physical properties of a galaxy from photometry over a large range of wavelengths. By modeling the underlying stellar population and its interaction with the galaxy's ISM, and fitting the observed photometry, estimates of the star formation history, metallicity, stellar mass, etc. can be retrieved from the SED alone \citep[for a review, see][and references therein]{conroyModelingPanchromaticSpectral2013}. As a result, representative templates of galaxies and AGNs can be generated that span the majority of their physical parameter space. By fitting these templates to galaxy photometry, the relative contribution of an AGN to the SED can be determined. 

\subsection{Template Fitting}

To fit the optical to mid-IR photometry of the WISE Matched SNR sample, we use a straightforward template-fitting prescription. As we are only concerned with the relative AGN contribution to the SED and the extinction of the AGN, and not the multitude of galaxy parameters possible, a template-fitting approach is ideal for reducing the number of free parameters in our fitting. We make use of the elliptical (E), spiral (Sbc), and irregular (IM) empirical galaxy templates from \citet{assefLowResolutionSpectralTemplates2010}. In addition, we make use of the `normal' AGN template from \citet{lyuDustdeficientPalomarGreenQuasars2017a} along with the AGN extinction curve from Lyu et al. (in prep). We perform two fits, one using only the three galaxy templates, and another using the three galaxy templates plus the AGN template and AGN extinction curve. Only the AGN component has an extinction curve applied, the galaxy templates are used unextincted. For each object, the templates are redshifted to the spectroscopic redshift of the galaxy and we fit all available SDSS and WISE photometry. We convolve the appropriate filter curves with the templates to generate the model photometry from the templates. The template is interpolated to the filter wavelength using the flux preserving \texttt{SpectRes} code \citep{carnallSpectResFastSpectral2017}.

To generate the initial guess to the fit, we first attempt to fit to the data with all templates unextincted using a linear-least-squares solution. If this results in a fit where all template coefficients are greater than zero, then the linear solution is used as the initial guess. Otherwise, we calculate the coefficients for each template such that each template makes an equal flux contribution in the \textit{W1} filter to be used as the initial guess. The photometry is then fit using a TRF algorithm using the initial guess where the extinction on the AGN component is allowed to vary, and the parameters are all constrained to be greater than or equal to zero.

To measure the probability distributions in the coefficients and extinction, we use a Markov chain Monte-Carlo approach. By iterating 50,000 times, we are able to generate the posteriors of the coefficients and extinction in addition to determining the set of coefficients that return the highest likelihood, i.e. lowest $\chi^2$. We again compare to an F-distribution to generate the probability that the addition of AGN template to the galaxy templates provides a statistically better fit.

To fit the SEDs of the galaxies in the WISE Matched SNR sample, we make again use of the HPC clusters. With an average runtime of 1:39\,min per object, we used $\sim$12,000 CPU hours on the HPC Ocelote cluster to fit the SEDs of the 416,288 galaxies in the WISE SNR matched sample.

\subsection{Example SED Fits}

We demonstrate the efficacy of our template fitting and F-value comparison method by comparing two objects in our photometric sample. In Figure \ref{fig:photexample}, we show results of SED modeling for SDSS objID 1237678617934102742 and 1237648674532295222, both identified as spectroscopic AGNs from our analysis. In the top panel for each object, we show the fit to the photometry with the galaxy templates and AGN template combined with an AGN extinction curve. In the bottom panel, we show the fit to the photometry with only the galaxy templates. Both objects are examples where an F-Test validated that the inclusion of an AGN component and extinction curve resulted in a statistically better fit, accounting for the increased degrees of freedom. SDSS objID 1237678617934102742 is an example of a unobscured AGNs, with an E($B-V$) consistent with zero. In this case, the galaxy templates cannot explain the high $f_{\rm \textit{W2}}/f_{\rm \textit{W1}}$ ratio, i.e. red \textit{W1$-$W2} color, as they underpredict the $u$, \textit{W1}, and \textit{W2} fluxes, and overpredict the \textit{W3}, and \textit{W4} fluxes. Similarly, the galaxy templates cannot explain the high $f_{\rm \textit{W2}}/f_{\rm \textit{W1}}$ ratio, i.e. red \textit{W1$-$W2} color, in SDSS objID 1237648674532295222. In trying to do so, the galaxy templates underpredict the \textit{W2}, \textit{W3}, and \textit{W4} fluxes. The inclusion of an obscured AGN component with an E($B-V)\sim7\pm1$ produces a more robust fit and is able to better fit the red \textit{W1$-$W2} along with the remaining WISE fluxes. 

These objects provide good examples of where our SED fitting technique succeeds in identifying evidence for the presence of both unobscured and obscured AGN from optical and mid-IR photometry. The \texttt{FITS} tables with our concatenated photometric results are available on Zenodo:\dataset[doi:10.5281/zenodo.5848048]{https://doi.org/10.5281/zenodo.5848048}. In Table \ref{tab:photores}, we describe the columns of the photometric results table.

\begin{figure*}[ht!]
    \includegraphics[width=\textwidth]{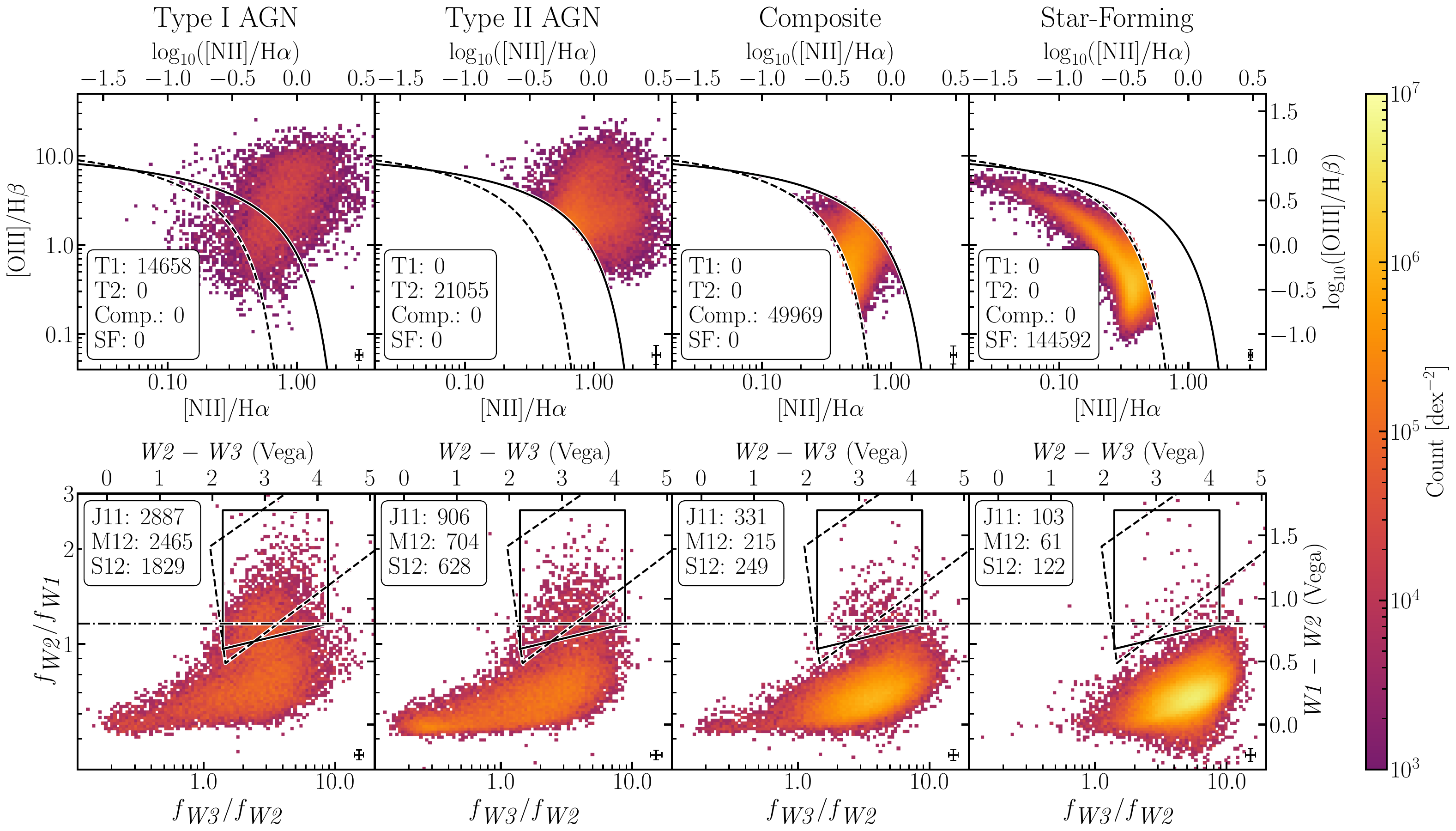}{}
    \caption{BPT (top) and WISE color-color (bottom) distributions for optical Type I AGN (T1; left), optical Type II AGN (T2; center left), BPT composite (Comp.; center right),  BPT star-forming (SF; right) for galaxies with a flux SNR $>3$ in the represented emission lines. We note that T1 galaxies without a flux SNR $>3$ in the BPT emission lines do not appear in the BPT diagram but are still counted in the totals. Typical WISE selection criteria are successful at excluding star-forming objects and preferentially select T1 objects, and AGNs are found throughout WISE color-color space. Refer to Figure \ref{fig:sample} for a description of the BPT delineations, WISE color-color selection criteria, and error bars plotted in black.\label{fig:BPTclass}}
\end{figure*}

\begin{figure*}[ht!]
    \centering
    \includegraphics[width=0.75\textwidth]{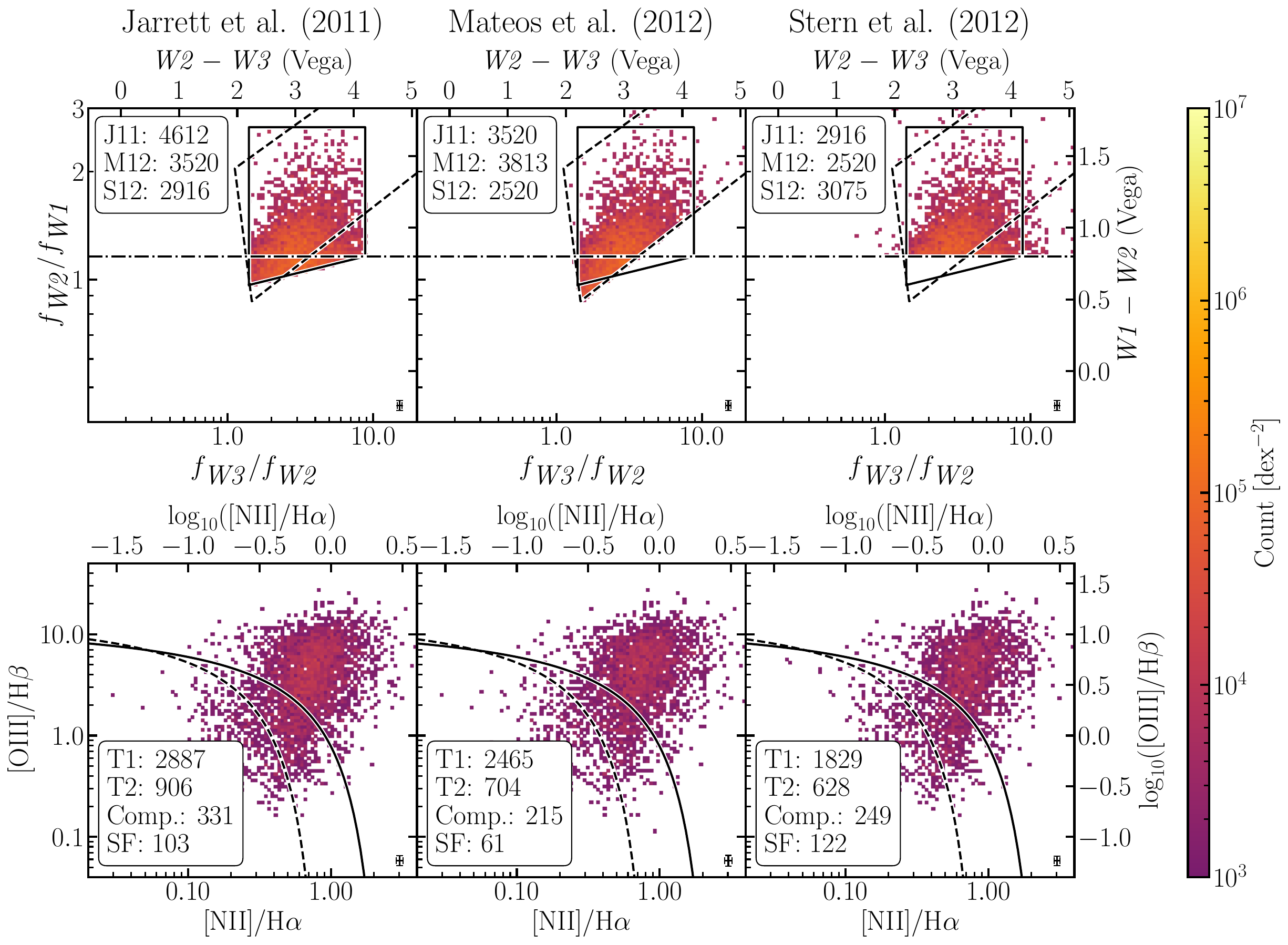}{}
    \caption{WISE color-color (top) and BPT (bottom) and distributions for \citet{jarrettSpitzerWISESurveyEcliptic2011} selected (J11; left), \citet{sternMidInfraredSelectionAGN2012} selected (S12; center), and \citet{mateosUsingBrightUltraHard2012} selected (M12; right) objects. We note that T1 galaxies without a flux SNR $>3$ in the represented lines do not appear in the BPT diagram but are still counted in the total. Typical WISE selection criteria are successful at excluding star-forming objects and preferentially select T1 objects. Refer to Figure \ref{fig:sample} for a description of the BPT delineations, WISE color-color selection criteria, and error bars plotted in black.\label{fig:WISEselect}}
\end{figure*}

\begin{figure}[ht!]
    \centering
    \includegraphics[width=\columnwidth]{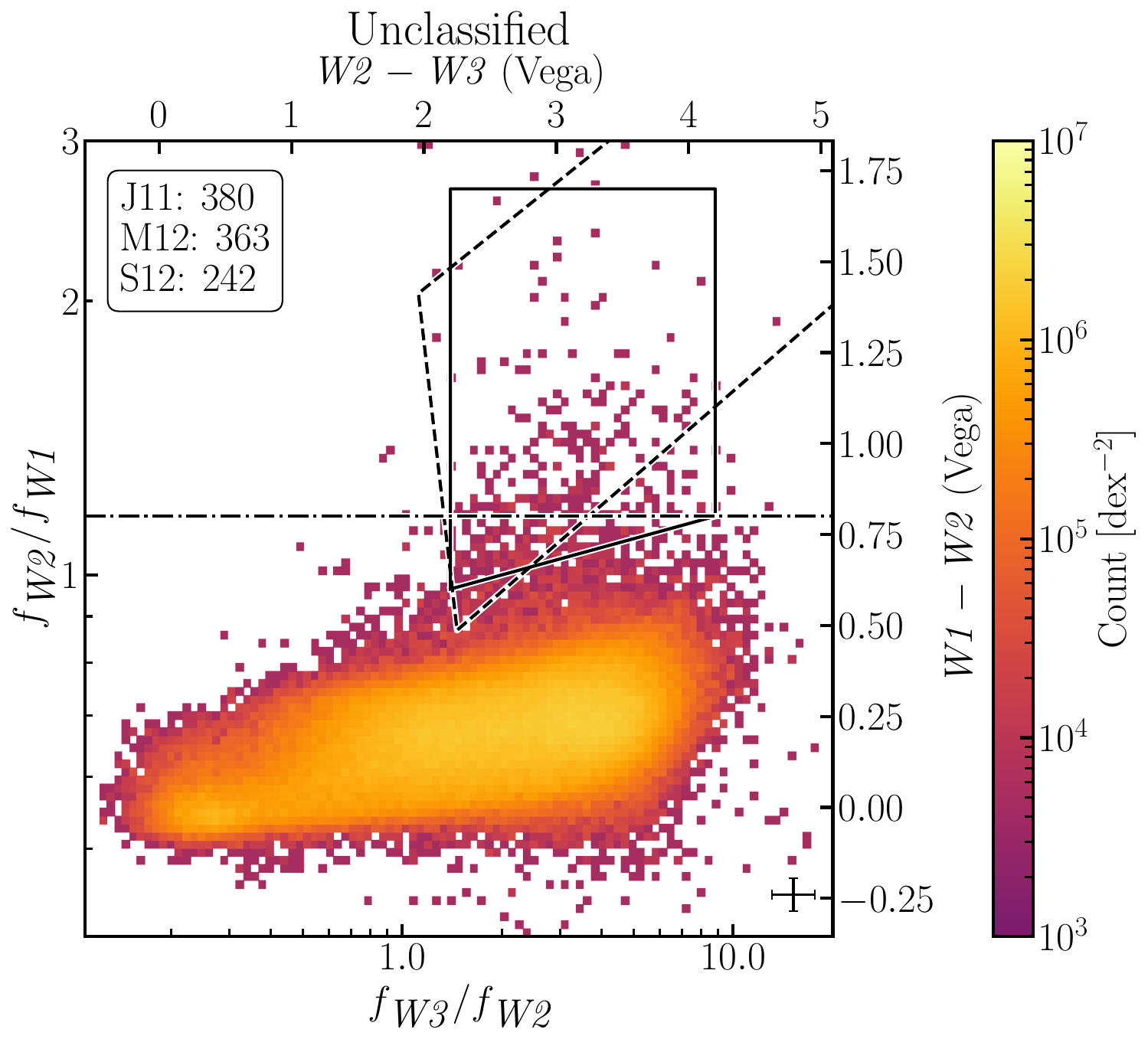}{}
    \caption{The WISE color-color distributions for the Unclassified classification. These objects are characterized by low $f_{\rm \textit{W2}}/f_{\rm \textit{W1}}$ ratios, i.e. blue \textit{W1$-$W2} colors, and most are outside of typical mid-IR AGN selection criteria (see Table \ref{tab:select}). Refer to Figure \ref{fig:sample} for a description of the BPT delineations, WISE color-color selection criteria, and error bars plotted in black.\label{fig:unclassified}}
\end{figure}

\section{Results}\label{sec:result}

To understand how optical spectroscopic AGNs are distributed throughout WISE color space, we combine the results of our spectroscopic fitting and SED analysis. In Figure \ref{fig:BPTclass}, we plot the BPT and WISE color-color distributions of spectroscopically classified objects. We show the distribution of T1, T2, Comp, and SF galaxies on the BPT diagnostic, and the corresponding distribution in WISE color space. BPT star-forming and composite galaxies tend to have high $f_{\rm \textit{W3}}/f_{\rm \textit{W2}}$ ratios while at relatively low  $f_{\rm \textit{W2}}/f_{\rm \textit{W1}}$ ratios, i.e. red \textit{W2$-$W3} colors while at blue \textit{W1$-$W2} colors. While these galaxies contain hot dust potentially due to star-formation, it is not heated to sufficient temperatures to result in higher $f_{\rm \textit{W2}}/f_{\rm \textit{W1}}$ ratios, i.e. redder \textit{W1$-$W2} color. There are relatively small number of BPT star-forming and composite galaxies (on the order of 10$^2$) that are present in the AGN selection regions. Optical Type I and II AGNs are ubiquitous throughout WISE color space and inhabit the same regions dominated by composite and star-forming galaxies. However, the presence of hot dust from AGN heating does push AGNs to higher $f_{\rm \textit{W2}}/f_{\rm \textit{W1}}$ ratios, i.e. redder \textit{W1$-$W2} colors, as expected. We find that typical WISE selection criteria are primarily successful at avoiding objects with star-forming and composite line ratios.

In Figure \ref{fig:WISEselect}, we explore where mid-IR selected galaxies are found in BPT space. WISE selection criteria predominantly select optical spectroscopic AGNs, optical Type I AGNs, and select relatively few star-forming galaxies. While the distributions of the WISE selection criteria in BPT space are similar, they differ mainly in the numbers of AGNs recovered. To compare different selection techniques, we define accuracy as the total number of objects in the target class, in this case optical Type I or Type II AGN, that are selected divided by the total number of objects selected by the color selection regardless of classification. We define completeness as the total number of objects in the target class that are selected divided by the total number of objects in the target class, selected or not. In Table \ref{tab:select} we tabulate the number of selected objects by BPT classification for each of the three mid-IR color selection criterion: the \citet{jarrettSpitzerWISESurveyEcliptic2011} box, the \citet{mateosUsingBrightUltraHard2012} wedge, and the \citet{sternMidInfraredSelectionAGN2012} line. On average they have an accuracy of $\sim$82\%, and recover around $\sim$7.5\% of all Type I and Type II AGN. 

However, a large portion of the WISE Matched SNR sample is made up of galaxies which we are unable to classify. Therefore, we can only place lower bounds on the accuracies of the different mid-IR color selection criterion based on the number of Unclassified galaxies left in the selection. In Figure \ref{fig:unclassified} we plot the distribution of objects in the Unclassified category. These objects are characterized by low $f_{\rm \textit{W2}}/f_{\rm \textit{W1}}$ ratios, i.e. blue \textit{W1$-$W2} colors, outside of typical mid-IR selection criteria. Less than 0.2\% of all Unclassified objects lie inside any mid-IR criterion and make up around only 10\% of the objects selected. While it is possible that these objects do indeed host an AGN that could be identified with deeper optical spectroscopy, given that relatively few of them are actually selected by mid-IR criteria, these objects do not affect the remainder of the analysis in this work.

\begin{figure}[ht!]
    \includegraphics[width=\columnwidth]{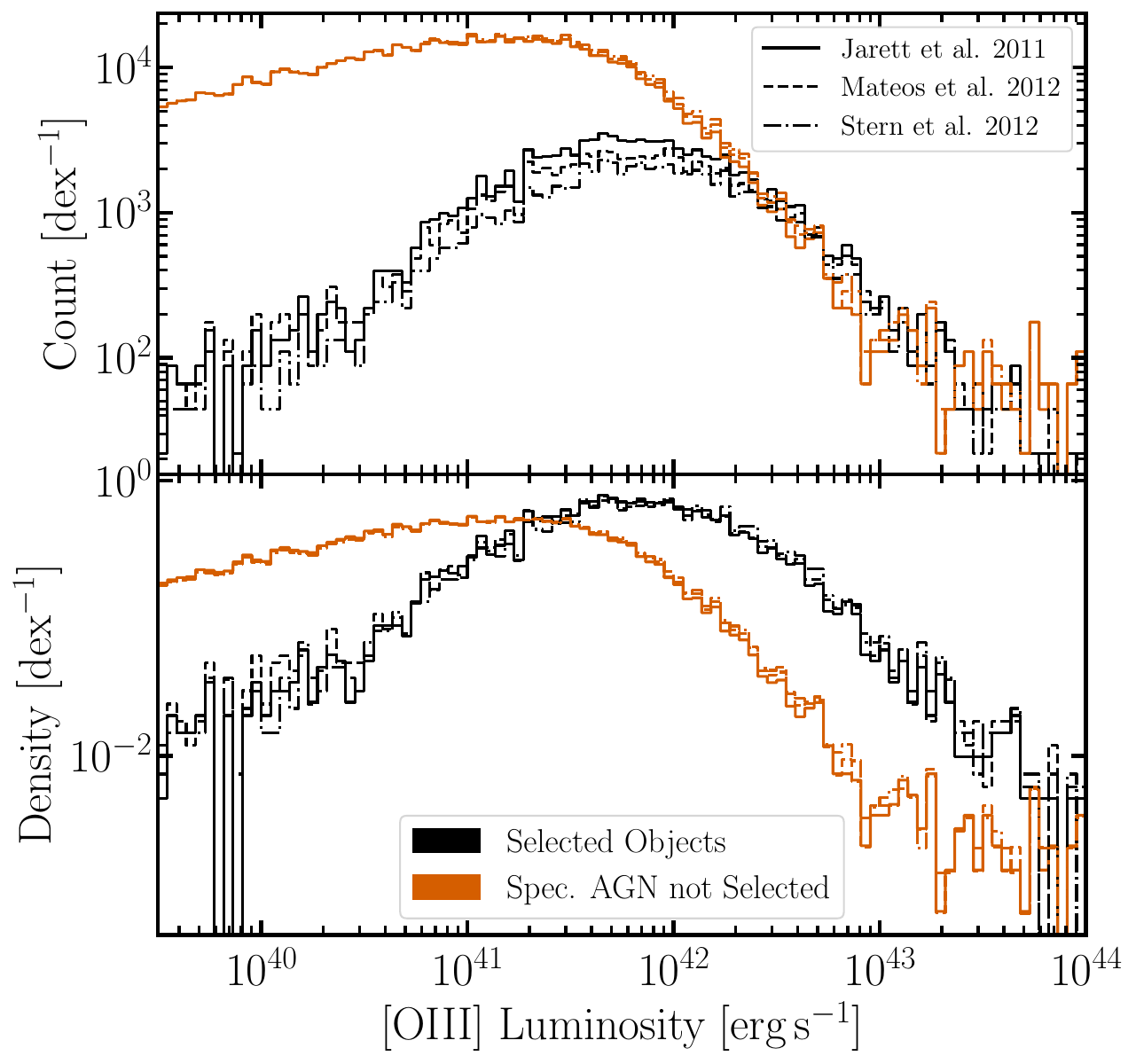}{}
    \caption{Distributions of [\ion{O}{3}]$\lambda$5007\AA\ luminosity for objects selected (black) and spectroscopic AGN not selected (orange) by different mid-IR color selection techniques. In the top panel we show the count histogram, while in the bottom panel we show the density histogram. The luminosity is corrected for galactic-scale extinction using the measured Balmer decrement. Typical mid-IR selection criteria are biased towards selecting a more luminous AGN population.\label{fig:OIII}}
\end{figure}

To understand what kind of objects are selected by the \citet{jarrettSpitzerWISESurveyEcliptic2011}, \citet{mateosUsingBrightUltraHard2012}, and \citet{sternMidInfraredSelectionAGN2012} selection criteria, in Figure \ref{fig:OIII}, we plot [\ion{O}{3}] luminosity distributions for each of these selections along with the spectroscopic AGNs that are not selected, i.e. exist outside each of the mid-IR selection criteria. [\ion{O}{3}]$\lambda$5007\AA\ is a good indicator for the luminosity of the AGN \citep[e.g.][]{bassaniThreedimensionalDiagnosticDiagram1999,heckmanRelationshipHardXRay2005,lamassaIndicatorsIntrinsicActive2010}. We apply a reddening correction by using the observed Balmer ratio (H$\alpha$/H$\beta$) compared to an intrinsic value of 3.1 combined with the \citet{fitzpatrickCorrectingEffectsInterstellar1999} extinction template (assuming an $R_V = 3.1$) to mitigate the effects of galactic-scale extinction. We find that typical mid-IR selection criterion are biased towards more luminous [\ion{O}{3}], i.e. more luminous AGNs. As AGN output decreases, it becomes increasingly difficult to discern the AGN from the host galaxy light from photometry alone. However, it may be possible to extend the current mid-IR selection criterion for an optical matched mid-IR sample to select a more complete sample of optical spectroscopic AGNs that probe a lower intrinsic AGN luminosity.

\begin{deluxetable*}{l|rr|rr|rr|rr|rr}
    \tablecaption{WISE Color Selection Criterion Statistics\label{tab:select}}
    \tablehead{
        \multicolumn{1}{c|}{} &
        \multicolumn{2}{c|}{J11} &
        \multicolumn{2}{c|}{M12} &
        \multicolumn{2}{c|}{S12} &
        \multicolumn{2}{c|}{This Work} &
        \multicolumn{2}{c}{This Work + Comp.}
    }
    \startdata
    Type I AGN      & 2887 & (62.6\%) & 2465 & (64.6\%) & 1829 & (59.5\%) & 4129 & (59.3\%) & 4655 & (55.1\%) \\
    \hline
    Type II AGN     & 906 & (19.6\%) & 704 & (18.5\%) & 628 & (20.4\%) & 1410 & (20.3\%) & 1728 & (20.5\%) \\
    \hline
    BPT Comp.   & 331 & (7.2\%) & 215 & (5.6\%) & 249 & (8.1\%) & 520 & (7.5\%) & 797 & (9.4\%) \\
    \hline
    BPT SF          & 103 & (2.2\%) & 61 & (1.6\%) & 122 & (4.0\%) & 133 & (1.9\%) & 279 & (3.3\%) \\
    \hline
    Stars           & 5 & (0.1\%) & 5 & (0.1\%) & 5 & (0.2\%) & 5 & (0.1\%) & 5 & (0.1\%) \\
    \hline
    Unclassified    & 380 & (8.2\%) & 363 & (9.5\%) & 242 & (7.9\%) & 762 & (10.9\%) & 983 & (11.6\%) \\
    \hline
    \multicolumn{1}{l|}{Total Selected} & \multicolumn{2}{c|}{4612} & \multicolumn{2}{c|}{3813} & \multicolumn{2}{c|}{3075} & \multicolumn{2}{c|}{6959} & \multicolumn{2}{c|}{8447} \\
    \hline
    \hline
    \multicolumn{1}{c|}{AGN Accuracy} & \multicolumn{2}{c|}{82.2\%} & \multicolumn{2}{c|}{83.1\%} & \multicolumn{2}{c|}{79.9\%} & \multicolumn{2}{c|}{79.6\%} & \multicolumn{2}{c|}{75.6\%} \\
    \multicolumn{1}{c|}{AGN + Comp. Accuracy} & \multicolumn{2}{c|}{89.4\%} & \multicolumn{2}{c|}{88.7\%} & \multicolumn{2}{c|}{88.0\%} & \multicolumn{2}{c|}{87.1\%} & \multicolumn{2}{c|}{85.0\%} \\
    \hline
    \multicolumn{1}{c|}{AGN Completeness} & \multicolumn{2}{c|}{10.6\%} & \multicolumn{2}{c|}{8.9\%} & \multicolumn{2}{c|}{6.9\%} & \multicolumn{2}{c|}{15.5\%} & \multicolumn{2}{c|}{17.9\%} \\
    \multicolumn{1}{c|}{AGN + Comp. Completeness} & \multicolumn{2}{c|}{4.8\%} & \multicolumn{2}{c|}{3.9\%} & \multicolumn{2}{c|}{3.2\%} & \multicolumn{2}{c|}{7.1\%} & \multicolumn{2}{c|}{8.4\%} \\
    \enddata
    \tablecomments{Table containing the relative number of galaxies in each BPT class as selected by each WISE selection criterion: \citet[J11]{jarrettSpitzerWISESurveyEcliptic2011}, \citet[M12]{mateosUsingBrightUltraHard2012}, \citet[S12]{sternMidInfraredSelectionAGN2012}, and selection criteria presented in Section \ref{subsec:criterion} and Appendix \ref{sec:bptcomp}. In addition, we compute the relative accuracy and completeness, defined in Section \ref{subsec:criterion}, of each metric for either selecting optical Type I or Type II AGNs or selecting AGNs and composite galaxies, as described in Appendix \ref{sec:bptcomp}. As discussed in Section \ref{sec:result}, the accuracy and completeness statistics presented should be treated as lower limits due to the fraction of objects for which we cannot assign a spectroscopic classification.}
\end{deluxetable*}

\begin{figure*}[ht!]
    \includegraphics[width=\textwidth]{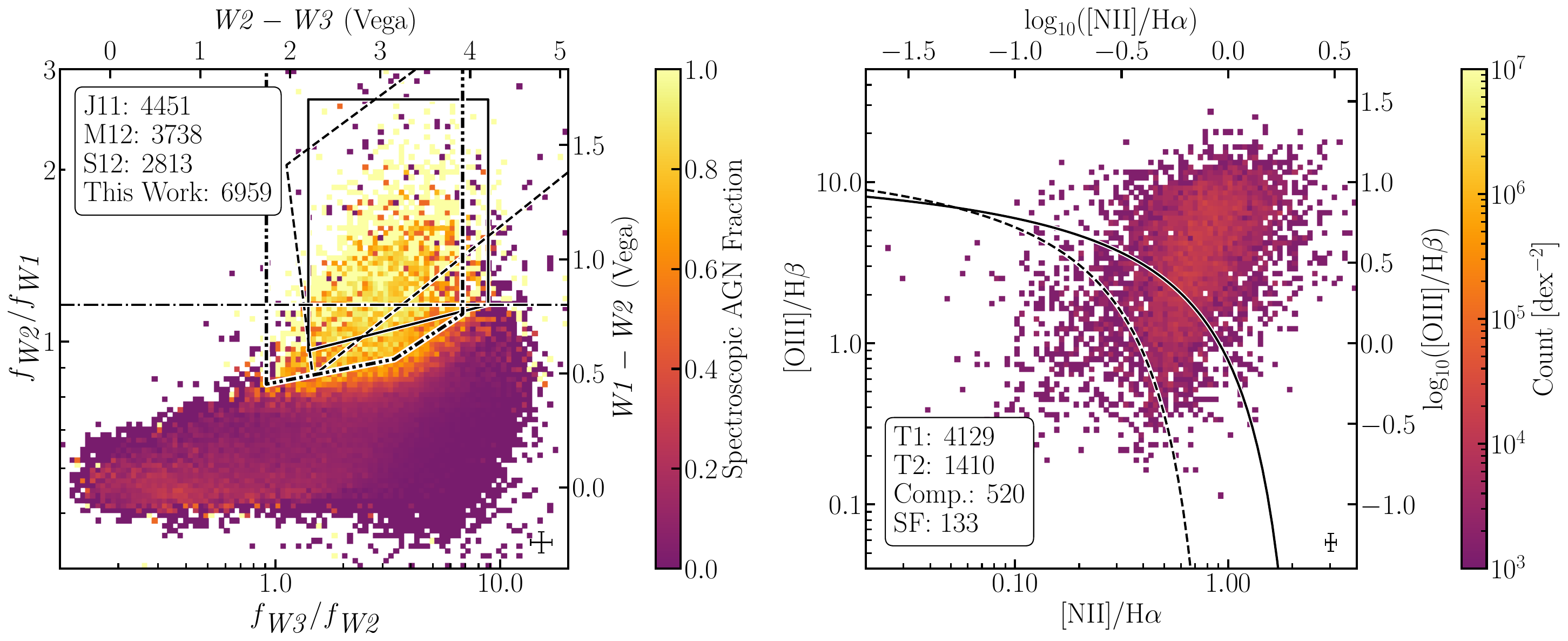}{}
    \caption{On the left we plot the fraction of spectroscopic AGNs (both Type I and Type II) as a function of WISE mid-IR color space. We present our new color selection as a dot-dot-dashed line, designed to target spectroscopic AGNs. On the right, we plot the distribution of where our selected objects are found in BPT space for those objects with detected emission lines. Refer to Figure \ref{fig:sample} for a description of the BPT delineations, WISE color-color selection criteria, and error bars plotted in black.\label{fig:Cut}}
\end{figure*}

\subsection{A New mid-IR AGN Criterion}\label{subsec:criterion}

Using the WISE Matched SNR sample, we search for a new selection criterion in WISE color-color space with the goal of selecting a more complete sample of optical spectroscopic AGN at comparable accuracies to previous mid-IR color selection. We parameterize our selection criterion by a minimum and maximum \textit{W2$-$W3} along with two linear cuts through WISE color-color space. To generate the selection criterion, we minimize an objective function, $f(\vec{x},t)$, presented in Equation \ref{eq:objective}, where $\vec{x}$ represents the parametrization of the selection criterion, $t$ represents the targeted accuracy, $A(\vec{x})$ is the accuracy of the selection criterion, and $C(\vec{x})$ is the completeness of the selection criterion. We design the objective function to achieve the target accuracy while simultaneously maximizing the completeness of the selection. We estimate initial values by eye and minimize the objective function using the Nelder-Mead algorithm with a target accuracy of 80\%. Following the minimization, we retrieve a selection criterion defined by the three WISE flux ration inequalities in Equation \ref{eq:selectionflux} and their equivalents in WISE color inequalities in Equation \ref{eq:selectionmag}.

\begin{equation}\label{eq:objective}
    f(\vec{x},t) = \frac{1 + |{A(\vec{x})-t|}}{1 + C(\vec{x})}
\end{equation}

\begin{subequations}\label{eq:selectionflux}
    \begin{equation}
        0.911 < \frac{f_{\rm \textit{W3}}}{f_{\rm \textit{W2}}} < 6.795
    \end{equation}
    \begin{equation}
        \frac{f_{\rm \textit{W2}}}{f_{\rm \textit{W1}}} > 0.848\left(\frac{f_{\rm \textit{W3}}}{f_{\rm \textit{W2}}}\right)^{0.0771}
    \end{equation}
    \begin{equation}
        \frac{f_{\rm \textit{W2}}}{f_{\rm \textit{W1}}} > 0.678\left(\frac{f_{\rm \textit{W3}}}{f_{\rm \textit{W2}}}\right)^{0.261}
    \end{equation}
\end{subequations}

\begin{subequations}\label{eq:selectionmag}
    \begin{equation}
        1.734 < (\textrm{\textit{W2$-$W3}}) < 3.916
    \end{equation}
    \begin{equation}
        (\textrm{\textit{W1$-$W2}}) > 0.0771(\textrm{\textit{W2$-$W3}}) + 0.319
    \end{equation}
    \begin{equation}
        (\textrm{\textit{W1$-$W2}}) > 0.261(\textrm{\textit{W2$-$W3}}) - 0.260
    \end{equation}
\end{subequations}

Our criterion is presented in Figure \ref{fig:Cut} which shows the spectroscopic AGN fraction in the WISE color-color space, where a spectroscopic AGN is defined as being classified as an optical Type I or Type II AGN. We also plot the recovered BPT distribution of objects using this new selection criterion. Our selection recovers optical spectroscopic AGNs with an accuracy of 79.6\% and a completeness of 15.5\%. With an on average less than five percentage point drop in accuracy, we are able to achieve a $\sim$50\% increase in completeness compared to the other typical WISE selection criteria studied in this work. We are only selecting around one third of optical spectroscopic AGNs, but due to the contamination from host galaxy light in low-luminosity AGNs, it remains difficult to increase the recovery further through optical and mid-IR photometric selection alone. Our WISE selection criteria can be applied to optical photometry matched to WISE mid-IR photometry, for those objects with SDSS $r<17.77$\,mag, or equivalent.

\begin{figure}[ht!]
    \includegraphics[width=\columnwidth]{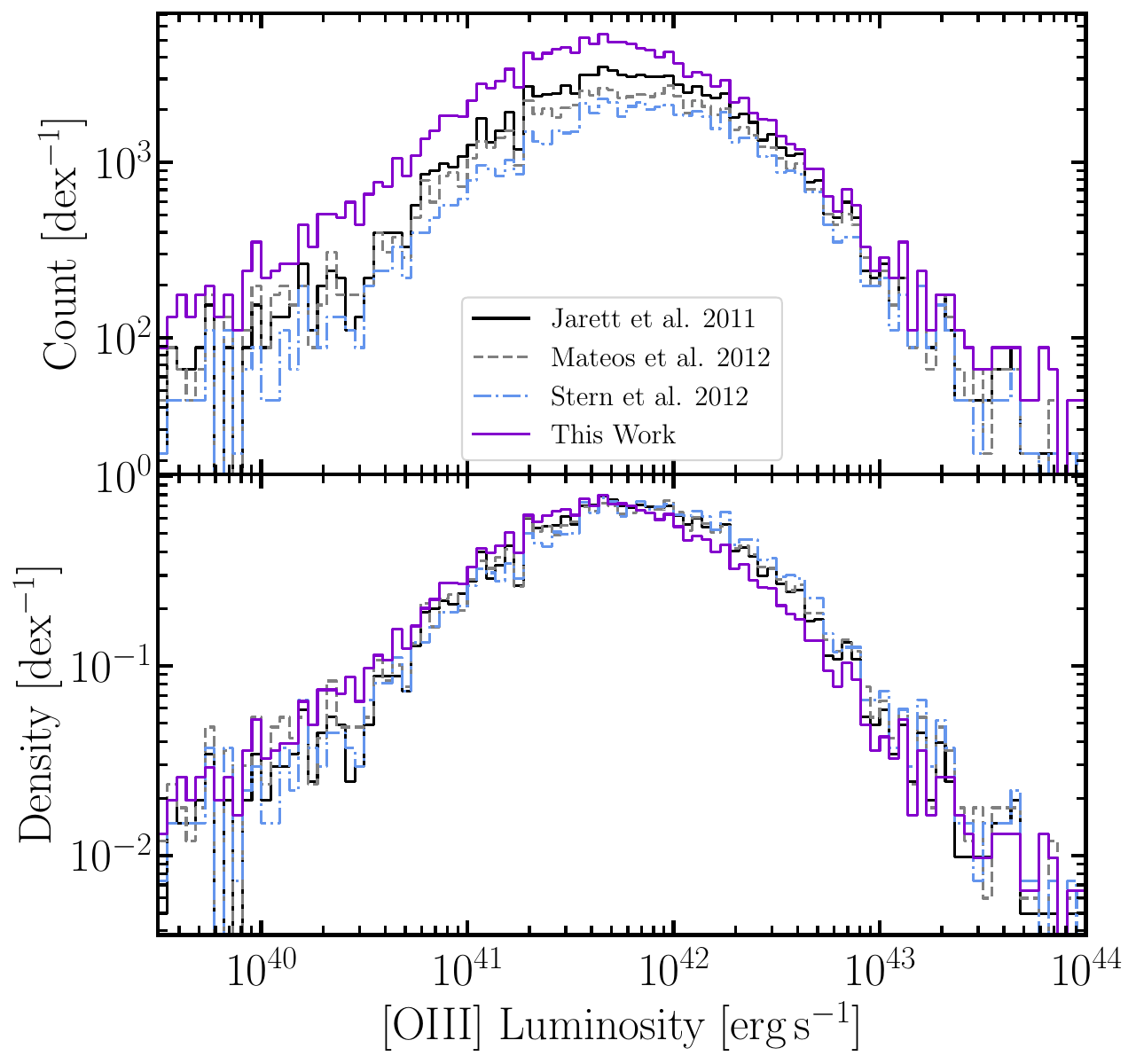}{}
    \caption{Distributions of [\ion{O}{3}]$\lambda$5007\AA\ luminosity for different mid-IR color selection techniques. In the top panel we show the count histogram, while in the bottom panel we show the density histogram. The luminosity is corrected for galactic-scale extinction using the measured Balmer decrement. The selection criteria presented in this work increases the completeness at all luminosities, but more notably at lower luminosities. This results in a median luminosity that is $>0.1$\,dex fainter than the other selection methods.\label{fig:oiii}}
\end{figure}

To understand how our selection differs from the \citet{jarrettSpitzerWISESurveyEcliptic2011}, \citet{mateosUsingBrightUltraHard2012}, and \citet{sternMidInfraredSelectionAGN2012} selection criteria, we compare the distributions of the spectroscopic and photometric properties derived from the WISE Matched SNR sample selected by the different selection methods. In Figure \ref{fig:oiii}, we investigate the distributions of the [\ion{O}{3}]$\lambda$5007\AA\ luminosity of objects of different mid-IR selection criteria. The luminosity is corrected for galactic-scale extinction using the measured Balmer decrement. Our selection criterion recovers nearly all of the objects selected by typical mid-IR selection criteria, but recovers an additional population of AGN at lower [\ion{O}{3}] luminosities. Our selected objects have, on average, $>0.1$\,dex lower median [\ion{O}{3}] luminosities than the other selection criteria. This reinforces that our selection criteria is indeed able to probe a lower luminosity regime than the other mid-IR selection criteria. 

To confirm whether the difference in [\ion{O}{3}] luminosities is driven by AGN luminosity and not another variable used in the calculation of the [\ion{O}{3}] luminosity, we investigate the distributions of the Balmer decrement (H$\alpha$/H$\beta$) and redshift. We find no significant differences in the Balmer decrement distributions ($<0.006$\,dex median difference) or redshift distributions ($<0.005$\,dex median difference) between the sample derived from our selection criteria and those of other typical WISE selections. Therefore, the observed difference in the observed [\ion{O}{3}] luminosity must be from our selection criterion probing an intrinsically less luminous population. In addition, we can conclude that the level of galactic-scale extinction in our selected sample likely does not differ from that of the other mid-IR selected samples.

\begin{figure}[ht!]
    \includegraphics[width=\columnwidth]{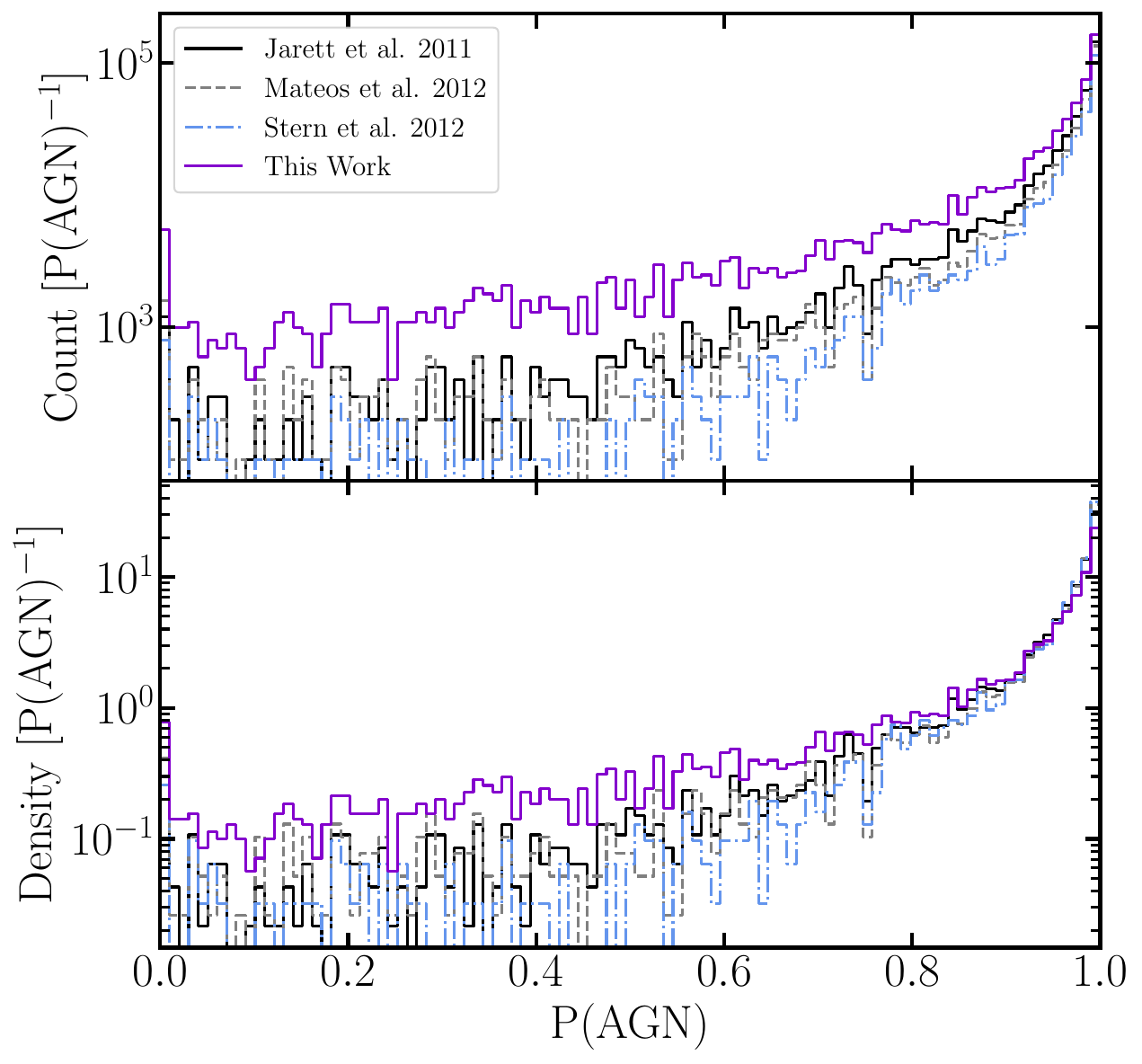}{}
    \caption{Distributions of the probability that an AGN component results in a statistically better fit as generated with an F-Test for different mid-IR color selection criteria. In the top panel we show the count histogram, while in the bottom panel we show the density histogram. The selection criterion presented in this work probes objects with a lower evidence for an AGN in their SED.\label{fig:FvalHist}}
\end{figure}

\begin{figure*}[ht!]
    \includegraphics[width=0.5\textwidth]{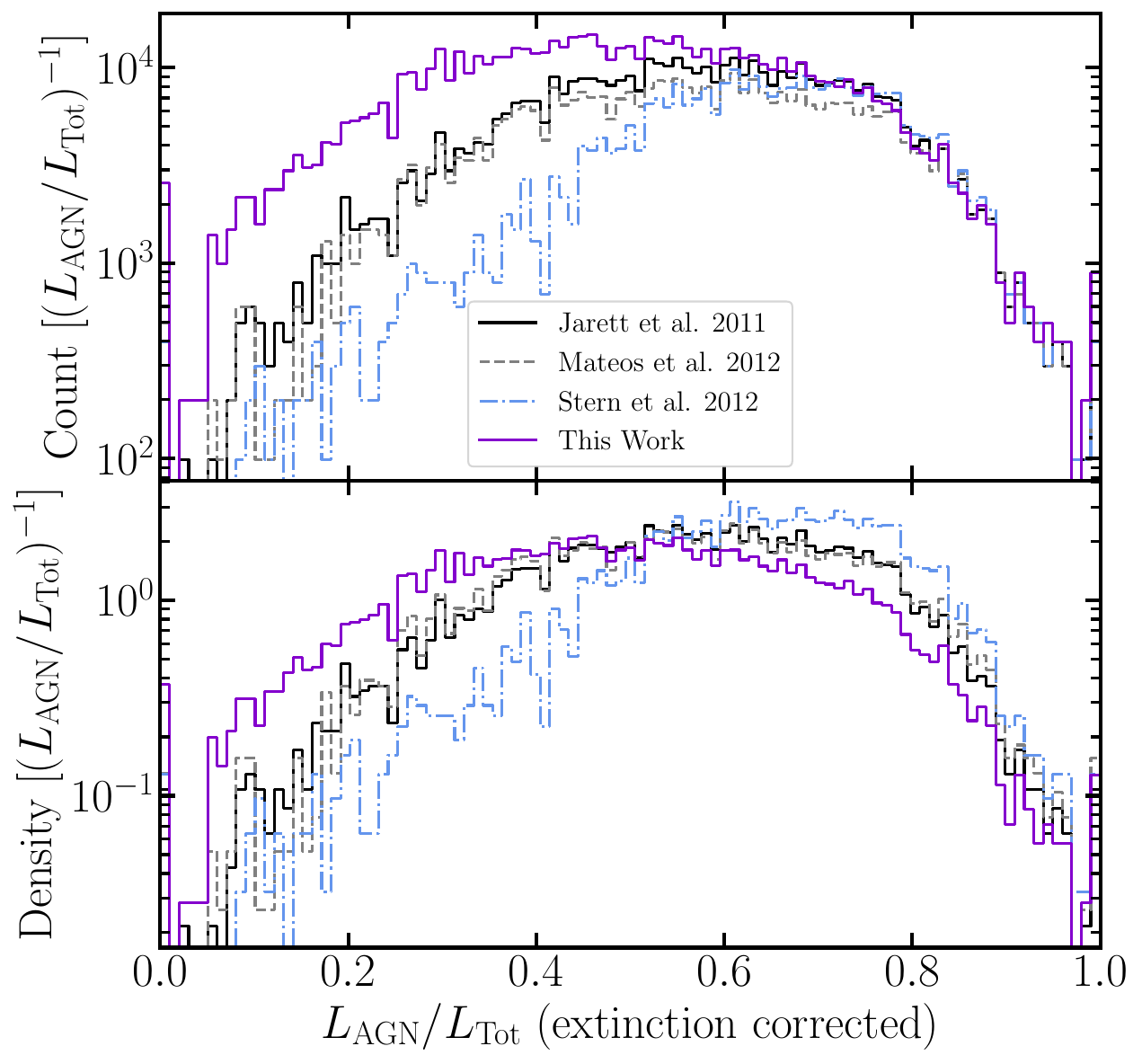}{}
    \includegraphics[width=0.5\textwidth]{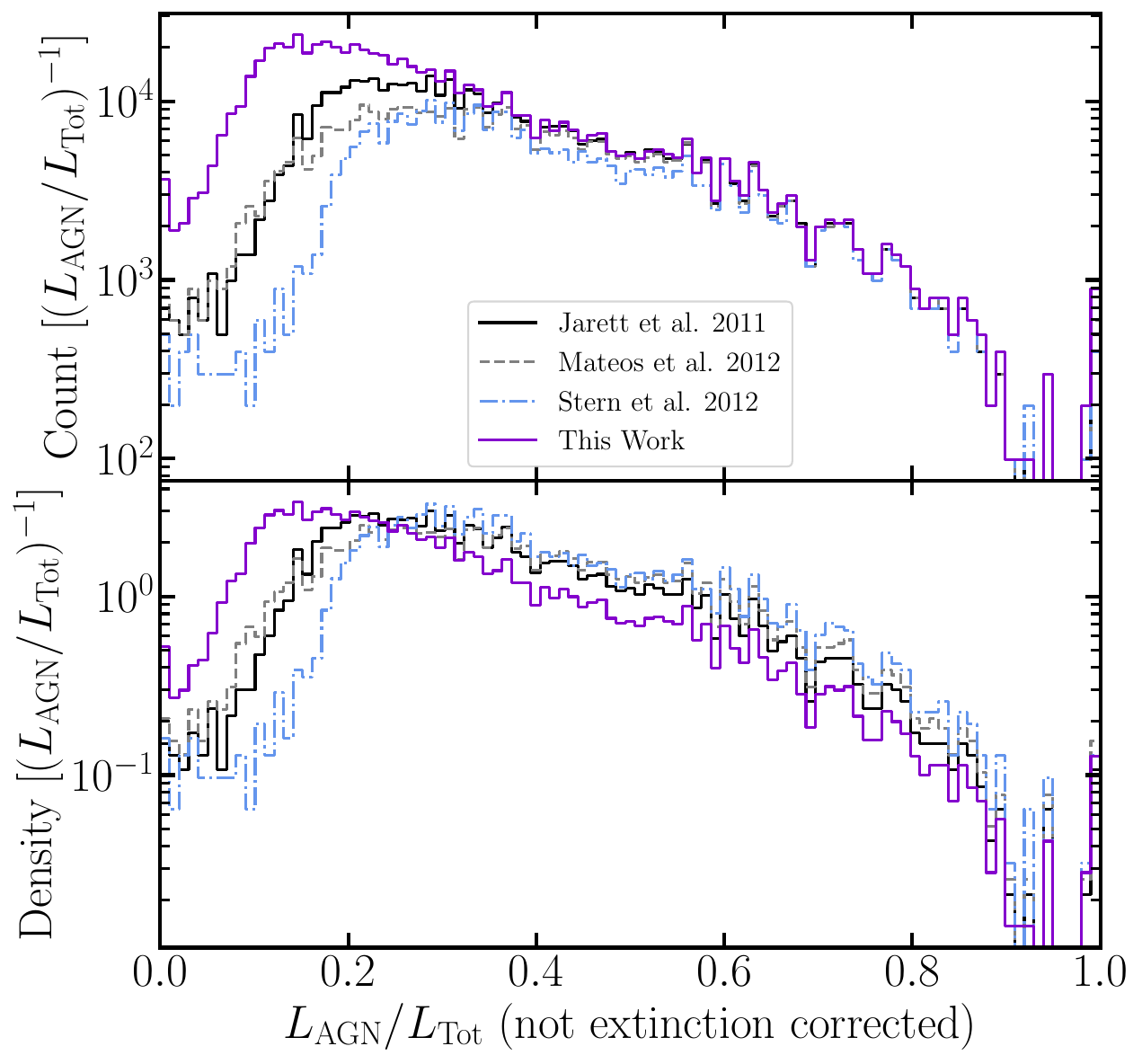}{}
    \caption{Distributions of $L_{\rm AGN}/L_{\rm Tot}$ for different mid-IR color selection criteria. On the left, the AGN luminosity is not corrected for extinction, while on the right it is corrected for extinction. In the top panels we show the count histograms, while in the bottom panels we show the density histograms. The selection criterion presented in this work probes objects with a lower AGN contribution to the SED.\label{fig:LfracHist}}
\end{figure*}

In Figure \ref{fig:FvalHist}, we plot the distributions of the probability that an AGN component is required for a statistically better fit as generated with an F-Test. We find that all mid-IR selection criterion typically select objects at high probability of requiring an AGN component, with tails out to lower probabilities. As other mid-IR selection were designed using templates and simulated colors, we should expect mid-IR selected samples to select objects that have a strong likelihood of requiring an AGN component. However, we find that our selection criterion selects more objects at a lower probability of requiring an AGN component than the other methods. As our new selection is probing a less luminous sample, we would expect that our objects would suffer from greater galaxy contamination and therefore be harder to detect in SED analysis. 

\begin{figure}[ht!]
    \includegraphics[width=\columnwidth]{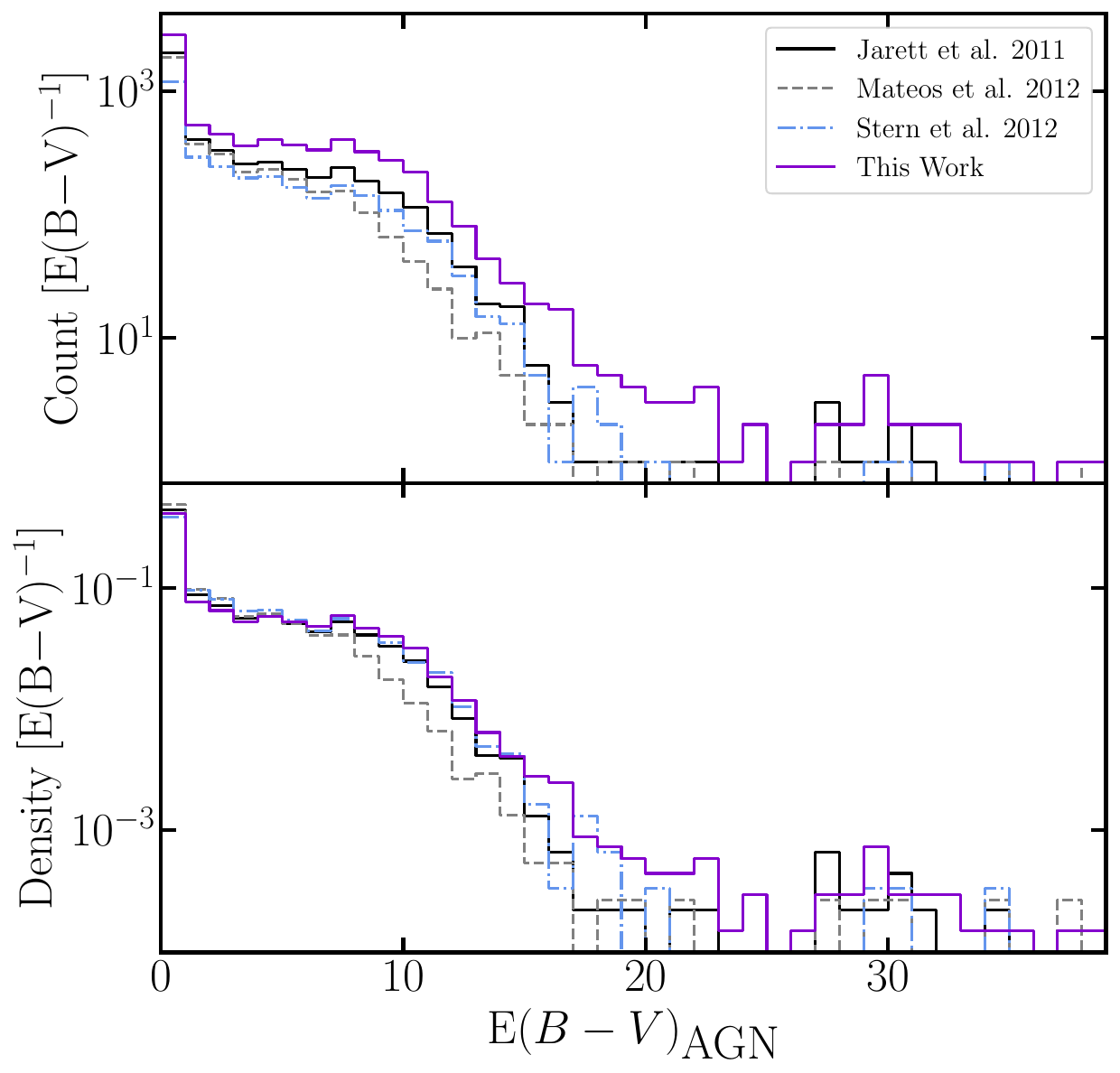}{}
    \caption{Distributions of E$(B-V)_\textrm{AGN}$ for different mid-IR color selection criteria. In the top panel we show the count histogram, while in the bottom panel we show the density histogram.\label{fig:EBVHist}}
\end{figure}

To confirm this hypothesis, we plot the distributions of the observed AGN luminosity contribution to the SED in the left panel of Figure \ref{fig:LfracHist}. The observed AGN luminosity contribution is, for the AGN fit, the integrated extincted AGN luminosity divided by the total luminosity of the SED from 0.1 to 30\,$\mu$m in the rest frame. As expected, we find that objects selected by our new selection criterion contribute less to the observed SED than the other mid-IR selected samples. We find that objects in our new selection have their AGN components contribute on average 20 percentage points less to the overall SED than the typical mid-IR selected samples. Even after correcting for the extinction, shown in the right panel of Figure \ref{fig:LfracHist}, this difference is still present, again strengthening the notion that our new selection is probing a less luminous set of AGN compared to their host galaxy than previous mid-IR selection criterion. 

Another way that AGN flux can be suppressed relative to the galaxy is an increase in the extinction, parameterized by E($B-V$), on the AGN. In Figure \ref{fig:EBVHist} we plot the distributions of the extinction of the AGN template in the fitting, E$(B-V)_\textrm{AGN}$. We find that the distributions are similar out to approximately an E$(B-V)_\textrm{AGN}=8$, after which our selected sample includes galaxies with slightly higher extinctions. This may indicate that we are probing a population that exhibits higher nuclear extinction. However, due to the degeneracies present in template fitting for lower luminosity AGNs, further study and follow up is required to confirm that this selection is indeed targeting a more heavily buried population of AGNs. 

In addition, it may be of interest to consider BPT composite objects as a target class as these objects may host AGNs. Immediately, the classification presented in this work returns target galaxies at comparable accuracies to other mid-IR selection criteria, e.g. at the 85-90\% level within a few percentage points and again see at a $\sim$50\% increase in completeness compared to the other selection criteria. In addition, we regenerate the selection criterion if we include composite galaxies as a target class. These results are presented in Appendix \ref{sec:bptcomp} and Table \ref{tab:select}. We find that including BPT composite galaxies generates a slightly relaxed color cut which, at approximately a 5\% decrease in accuracy, returns double the number of target galaxies.

\section{Conclusions and Discussion}\label{sec:conc}

In this work we have assembled a statistically representative sample of SDSS optical spectroscopy matched to mid-IR photometry from WISE to understand the distribution of spectroscopic AGNs in WISE color-color space. 

We present a newly developed spectroscopic fitting code, GELATO, with which we are able to measure the spectroscopic properties of the WISE Matched SNR sample and classify galaxies by their optical emission line ratios and additional emission line components. GELATO serves as an accurate tool for flexibly fitting rest-optical spectra where the relationship between many emission line species and additional components can be freely specified in a convenient framework to suit many scientific needs.

While AGNs are ubiquitous throughout WISE mid-IR color space, typical mid-IR selection succeeds in separating Type I and Type II AGNs from star-forming and composite galaxies, the high accuracy ($>80\%$) comes at the cost of low completeness ($\sim$9\%). In this work, we present a new selection criteria aimed at capturing a higher total number of Type I and Type II spectroscopic AGNs. This is the first mid-IR color selection defined by solely using the distribution of optical spectroscopic AGNs in WISE mid-IR color space. Our selection criterion is able to retrieve a sample of AGNs that is more than 50\% more complete than the aforementioned selections for WISE mid-IR photometry matched to optical photometry with $r<17.77$\,mag, with a completeness of over 13.8\%. In some cases, our selection criterion is nearly twice as complete as previous studies. Our selection criteria has an accuracy of 79.6\%, within 5 percentage points of the \citet{jarrettSpitzerWISESurveyEcliptic2011}, \cite{mateosUsingBrightUltraHard2012}, and \citet{sternMidInfraredSelectionAGN2012} selection criteria. In addition, if we include spectroscopic composite galaxies, we retrieve an additional mid-IR criterion that shows a similar relative increase in completeness at a marginal penalty for accuracy. 

While WISE photometry is available over the entire sky, the SDSS Legacy MGS is limited to a $<8000\deg^2$ region. At present, our selection criterion could be deployed to select AGNs from the Panoramic Survey Telescope and Rapid Response System \citep[Pan-STARRS;][]{chambersPanSTARRS1Surveys2016}, from the Hyper Suprime-Cam Subaru Strategic Program \citep{aiharaFirstDataRelease2018,aiharaSecondDataRelease2019}, and from the upcoming Legacy Survey of Space and Time (LSST) on the Vera C. Rubin Observatory. With new large area optical/near-IR photometric surveys developing rapidly in the near future without corresponding spectroscopy or deeper mid-IR surveys, our selection criteria can serve as a tool for selecting possible optical spectroscopic Type I or Type II AGNs at lower AGN luminosities and higher nuclear extinction than previous methods. 

In the future, we aim to extend this work down to fainter optical magnitudes, as upcoming surveys will continue to go deeper without corresponding improvements in all-sky mid-IR sensitivity. As we progress to fainter magnitudes, we probe higher redshifts, leading to evolution in color-color space, making the problem of finding AGNs without spectroscopic redshifts even more complicated. By using optical morphologies and advanced techniques including Machine Learning, we hope to search for new techniques to find spectroscopic AGNs using photometry alone. In addition, we aim to explore those objects without identifiable line emission in our spectroscopic sample, which may contain AGNs that are so heavily buried that there is no line emission at all.\\ 

REH acknowledges acknowledges support from the National Science Foundation Graduate Research Fellowship Program under Grant No. DGE-1746060.

KNH and MJR were supported by the National Aeronautics and Space Administration (NASA) Contract NAS50210 to the University of Arizona.

This material is based upon High Performance Computing (HPC) resources supported by the University of Arizona TRIF, UITS, and Research, Innovation, and Impact (RII) and maintained by the UArizona Research Technologies department.

We respectfully acknowledge the University of Arizona is on the land and territories of Indigenous peoples. Today, Arizona is home to 22 federally recognized tribes, with Tucson being home to the O’odham and the Yaqui. Committed to diversity and inclusion, the University strives to build sustainable relationships with sovereign Native Nations and Indigenous communities through education offerings, partnerships, and community service.  

This work makes use of color palettes created by Martin Krzywinski designed for colorblindness. The color palettes and more information can be found at \url{http://mkweb.bcgsc.ca/colorblind/}.

This publication makes use of data products from the Wide-field Infrared Survey Explorer, which is a joint project of the University of California, Los Angeles, and the Jet Propulsion Laboratory/California Institute of Technology, funded by the National Aeronautics and Space Administration.

Funding for the SDSS and SDSS-II has been provided by the Alfred P. Sloan Foundation, the Participating Institutions, the National Science Foundation, the U.S. Department of Energy, the National Aeronautics and Space Administration, the Japanese Monbukagakusho, the Max Planck Society, and the Higher Education Funding Council for England. The SDSS Web Site is http://www.sdss.org/.

The SDSS is managed by the Astrophysical Research Consortium for the Participating Institutions. The Participating Institutions are the American Museum of Natural History, Astrophysical Institute Potsdam, University of Basel, University of Cambridge, Case Western Reserve University, University of Chicago, Drexel University, Fermilab, the Institute for Advanced Study, the Japan Participation Group, Johns Hopkins University, the Joint Institute for Nuclear Astrophysics, the Kavli Institute for Particle Astrophysics and Cosmology, the Korean Scientist Group, the Chinese Academy of Sciences (LAMOST), Los Alamos National Laboratory, the Max-Planck-Institute for Astronomy (MPIA), the Max-Planck-Institute for Astrophysics (MPA), New Mexico State University, Ohio State University, University of Pittsburgh, University of Portsmouth, Princeton University, the United States Naval Observatory, and the University of Washington.

Funding for the Sloan Digital Sky Survey IV has been provided by the Alfred P. Sloan Foundation, the U.S. Department of Energy Office of Science, and the Participating Institutions. 


\software{Astropy \citep{collaborationAstropyCommunityPython2013}, GELATO \citep{hvidingTheSkyentistGELATOGELATO2022}, \LaTeX\ \citep{lamportLaTeXDocumentPreparation1994}, Matplotlib \citep{hunterMatplotlib2DGraphics2007}, NumPy \citep{oliphantGuideNumPy2006,vanderwaltNumPyArrayStructure2011, harrisArrayProgrammingNumPy2020}, SciPy \citep{virtanenSciPyFundamentalAlgorithms2020}, SpectRes \citep{carnallSpectResFastSpectral2017}}

\appendix
\restartappendixnumbering

\section{BPT Composite Galaxies}
\label{sec:bptcomp}

\begin{figure*}[ht!]
    \includegraphics[width=\textwidth]{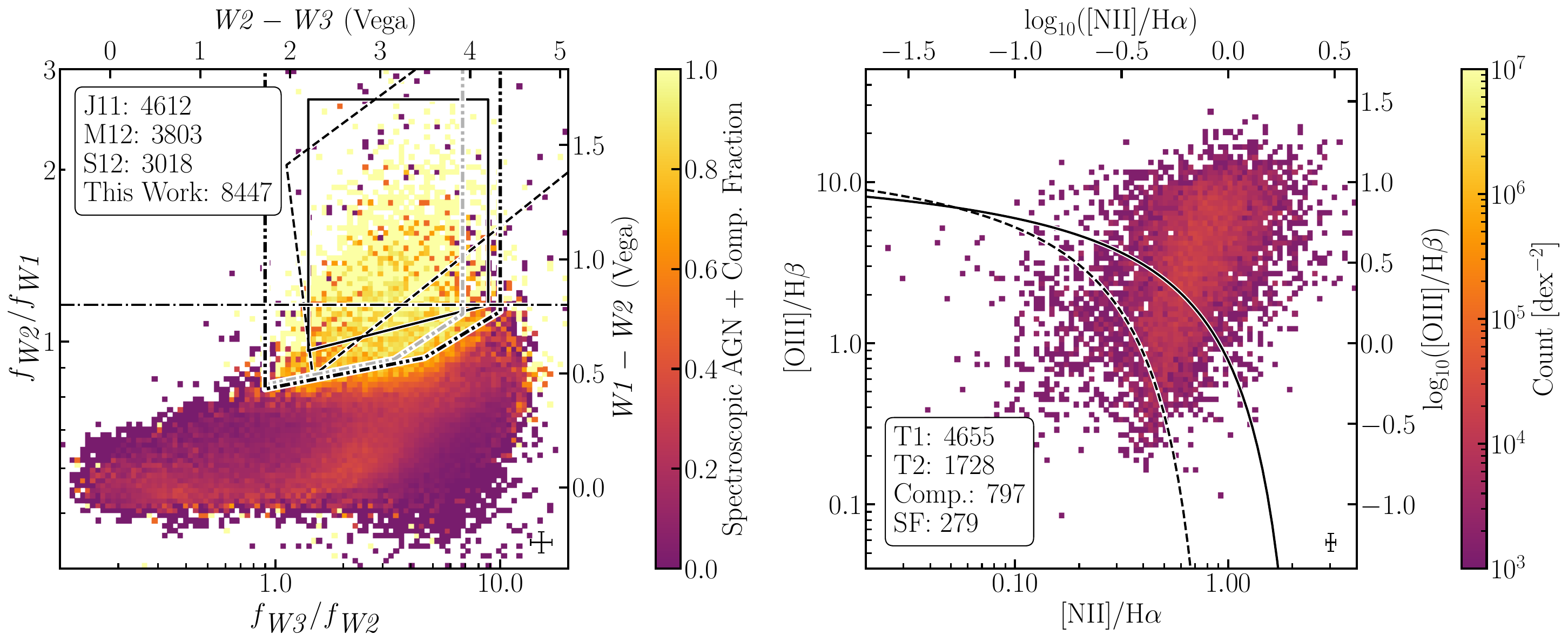}{}
    \caption{On the right we plot the fraction of spectroscopic AGNs (both Type I and Type II) and composite galaxies as a function of WISE mid-IR color space. We present the new color selection as a dot-dot-dashed line, designed to target spectroscopic AGNs and BPT composite galaxies. On the right, we plot the distribution of where our selected objects are found in BPT space for those objects with detected emission lines. Refer to Figure \ref{fig:sample} for a description of the BPT delineations, WISE color-color selection criteria, and error bars plotted in black. As a gray dot-dot-dashed line, we plot the color selection criterion from this work described in Section \ref{subsec:criterion}.\label{fig:CutComp}}
\end{figure*}

\begin{figure}[ht!]
    \includegraphics[width=0.5\textwidth]{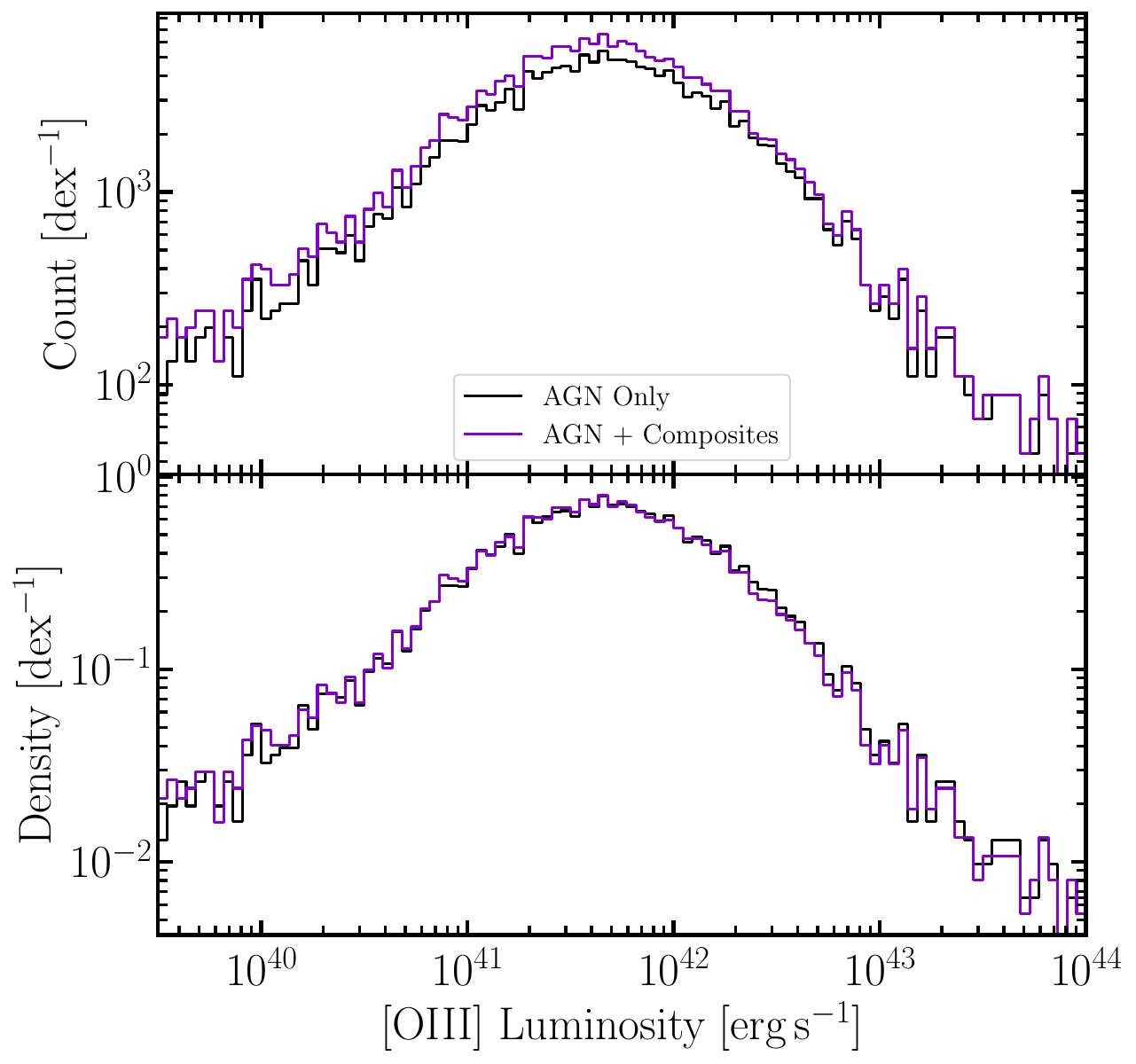}{}
    \includegraphics[width=0.5\textwidth]{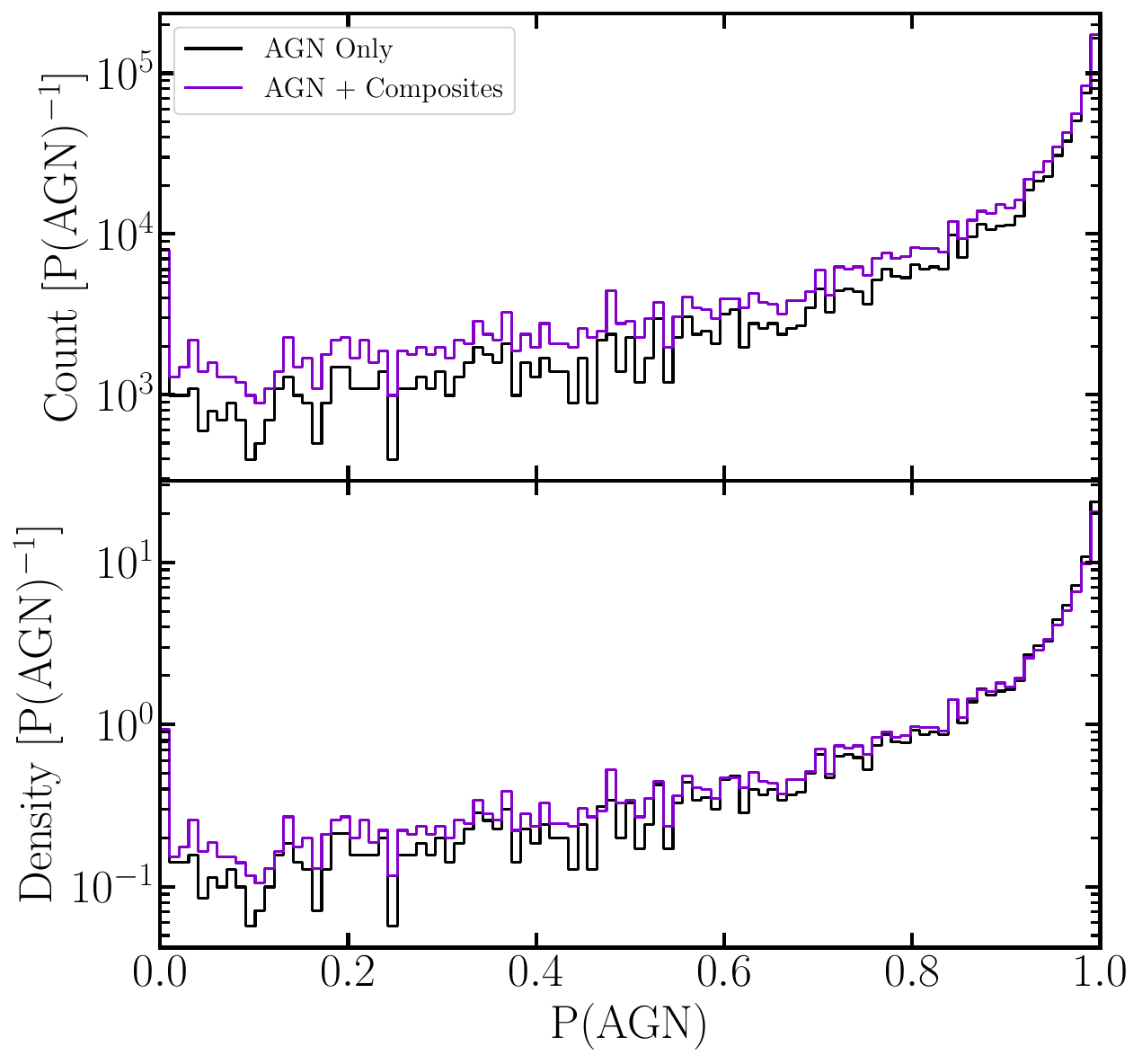}{}

    \caption{Distributions of [\ion{O}{3}] luminosity (left) and the probability that an AGN component results in a statistically better fit as generated with an F-Test (right) for different mid-IR color selection techniques. In the top panels we show the count histograms, while in the bottom panels we show the density histograms. The luminosity is corrected for galactic-scale extinction using the measured Balmer decrement. The selection criteria presented in this appendix increases the completeness at all luminosities, especially at lower luminosities, and probes galaxies with a lower evidence for an AGN in their SED.\label{fig:comp_oiii}}
\end{figure}

It is of interest to consider galaxies which lie within the composite region of the BPT diagram as these objects may be AGNs. This is especially relevant as BPT composite galaxies may contain lower luminosity AGN where the emission signature contaminated more strongly by star formation, or be comprised of interesting subclasses of Low Ionization Nuclear Emission Regions galaxies \citep{agostinoPhysicalDriversEmissionline2021}. In this section we investigate how drastically our selection criterion would change with the inclusion of BPT composite galaxies as a target class. We repeat the process outlined in Section \ref{subsec:criterion} to generate a new selection criterion for optical spectroscopic Type I AGNs, Type II AGNs, and composite galaxies. However, we now use a target accuracy of 85\% in line with the accuracies of other typical mid-IR color selection criteria if BPT composite galaxies are included. We present the defining three WISE flux ration inequalities in Equation \ref{eq:selectionfluxcomp} and their equivalents in WISE color inequalities in Equation \ref{eq:selectionmagcomp}. We plot the selection criterion in Figure \ref{fig:CutComp}, which also shows the spectroscopic AGN and composite galaxy fraction in the WISE color-color space along with the distribution of objects selected by the new selection criterion in the BPT diagram. We present the tabulated completeness and accuracy statistics for the composite selection criterion in Table \ref{tab:select}. We find that this selection criterion differs only slightly for the one trained on Type I and Type II AGN alone. Here we find that for a 5\% drop in accuracy, there is an over 70\% increase in completeness compared to the other mid-IR color selection criteria.

\begin{subequations}\label{eq:selectionfluxcomp}
    \begin{equation}
        0.899 < \frac{f_{\rm \textit{W3}}}{f_{\rm \textit{W2}}} < 9.990
    \end{equation}
    \begin{equation}
        \frac{f_{\rm \textit{W2}}}{f_{\rm \textit{W1}}} > 0.833\left(\frac{f_{\rm \textit{W3}}}{f_{\rm \textit{W2}}}\right)^{0.0771}
    \end{equation}
    \begin{equation}
        \frac{f_{\rm \textit{W2}}}{f_{\rm \textit{W1}}} > 0.648\left(\frac{f_{\rm \textit{W3}}}{f_{\rm \textit{W2}}}\right)^{0.241}
    \end{equation}
\end{subequations}

\begin{subequations}\label{eq:selectionmagcomp}
    \begin{equation} 
        1.720 < (\textrm{\textit{W2$-$W3}}) < 4.335
    \end{equation}
    \begin{equation}
        (\textrm{\textit{W1$-$W2}}) > 0.0771(\textrm{\textit{W2$-$W3}}) + 0.299
    \end{equation}
    \begin{equation}
        (\textrm{\textit{W1$-$W2}}) > 0.241(\textrm{\textit{W2$-$W3}}) - 0.275
    \end{equation}
\end{subequations}

\begin{figure*}[ht!]
    \includegraphics[width=0.5\textwidth]{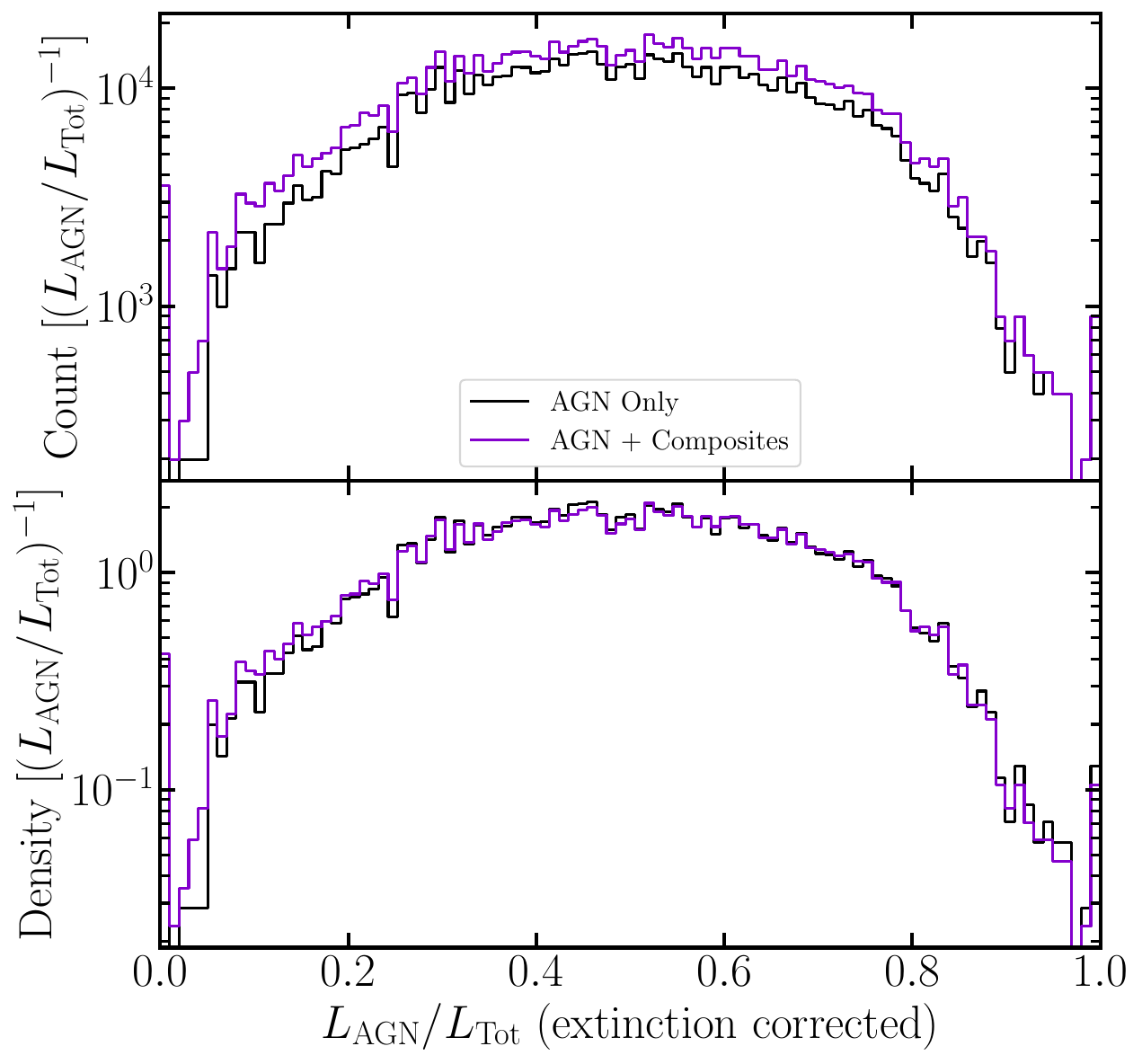}{}
    \includegraphics[width=0.5\textwidth]{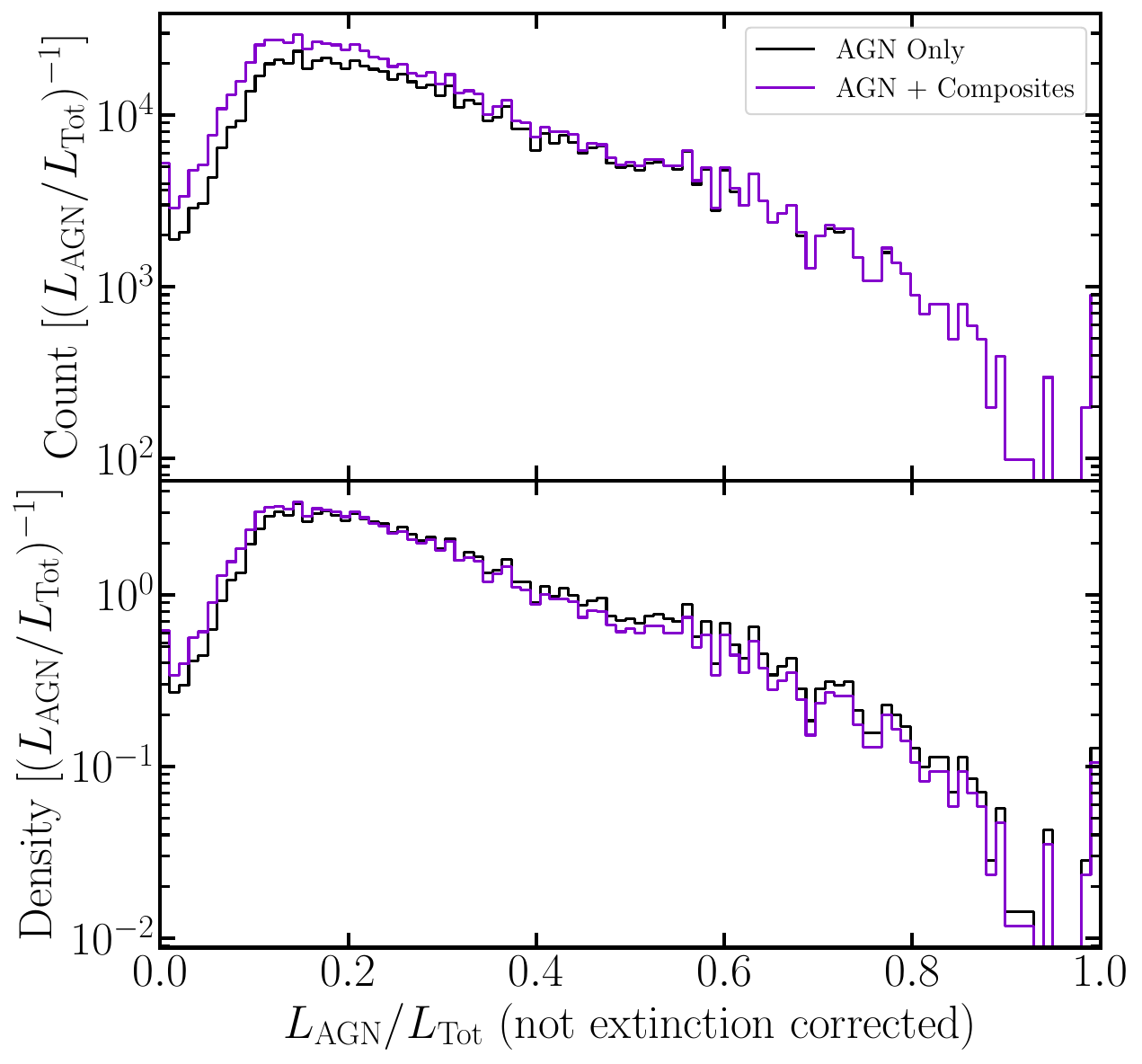}{}
    \caption{Distributions of $L_{\rm AGN}/L_{\rm Tot}$ for different mid-IR color selection criteria. On the left, the AGN luminosity is not corrected for extinction, while on the right it is corrected for extinction. In the top panels we show the count histograms, while in the bottom panels we show the density histograms. The selection criterion presented in this appendix probes objects with a lower AGN contribution to the SED.\label{fig:Comp_LfracHist}}
\end{figure*}

\begin{figure}[ht!]
    \center
    \includegraphics[width=0.5\textwidth]{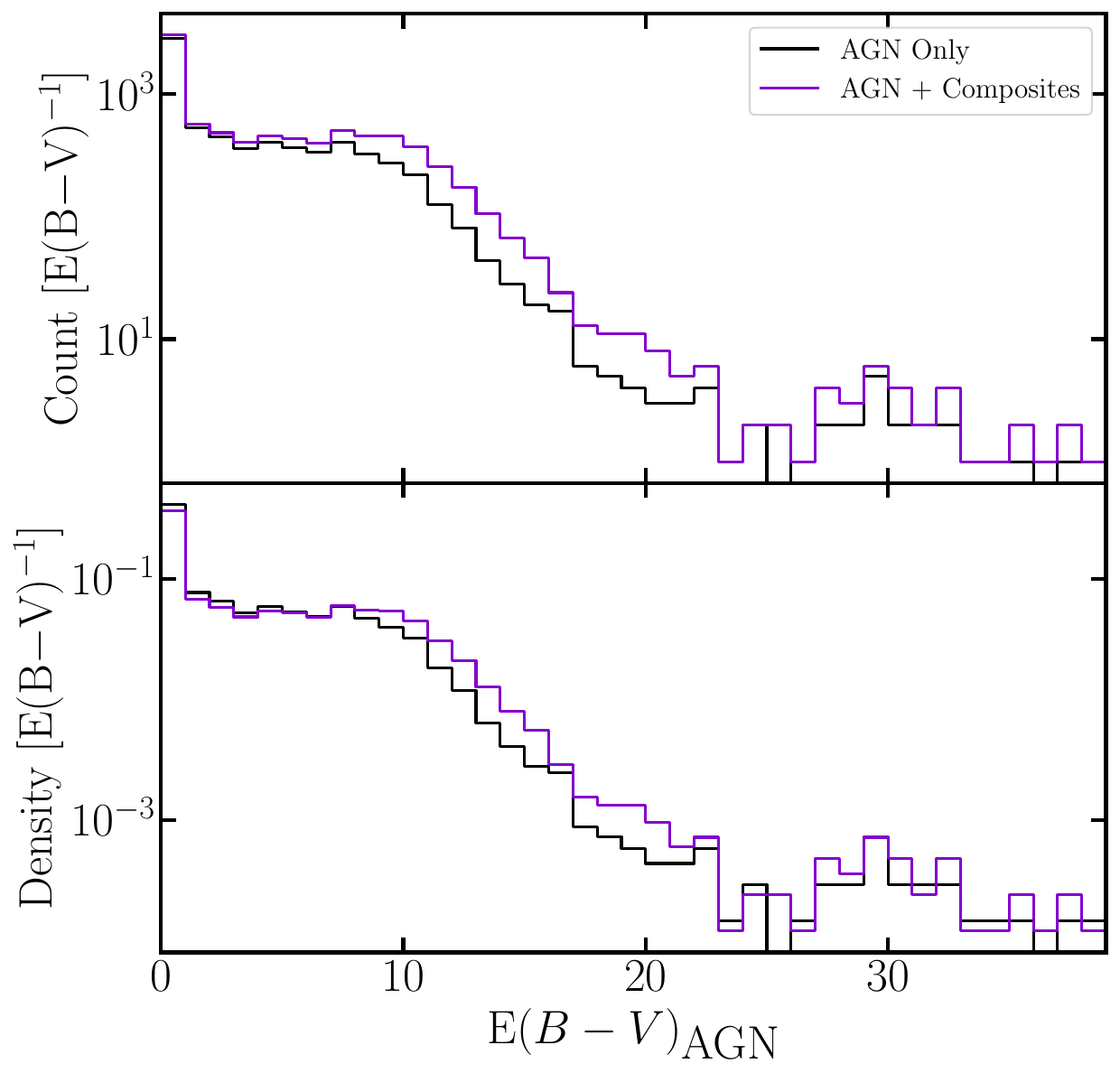}{}
    \caption{Distributions of E$(B-V)_\textrm{AGN}$ for different mid-IR color selection criteria. In the top panel we show the count histogram, while in the bottom panel we show the density histogram.\label{fig:Comp_EBVHist}}
\end{figure}

We compare the retrieved population from our selection of AGN and composite galaxies against our selection criterion presented in Section \ref{subsec:criterion} for AGN only. In Figure \ref{fig:comp_oiii} we present the distributions of the [\ion{O}{3}] luminosities and probabilities that the addition of an AGN component results in a statistically better fit. The selection criterion, now informed by both BPT AGN and composite galaxies, again returns a population at a lower [\ion{O}{3}] luminosity that have lower evidence for an AGN component from their SEDs, reinforcing the idea that we are probing a population at a lower overall luminosity. In addition, in Figure \ref{fig:Comp_LfracHist} we present distributions of the extinction uncorrected and corrected AGN contributions to the SED luminosity for this new selection criterion compared to typical mid-IR color selections. Once again, we find that our selection probes AGN at a lower luminosity relative to their host galaxy. Finally, in Figure \ref{fig:Comp_EBVHist}, we plot the distribution of the extinction on the AGN component, E$(B-V)_\textrm{AGN}$, this new selection criterion compared to typical mid-IR color selections. Even with the inclusion of composite galaxies, this approach to selection of AGN selects objects with evidence for higher levels of obscuration on the AGN from the SED. 

\section{Description of Data Tables}

\startlongtable
\begin{deluxetable*}{llcc}
\tablecaption{Description of WISE Matched SNR Sample Table Columns\label{tab:sample}}
\tablehead{
    \colhead{Name} &
    \colhead{Description} &
    \colhead{Unit} &
    \colhead{Type}
}
\startdata
sdss\_id & Unique SDSS objID & & int64 \\
sdss\_spec & Unique spectroscopic identifier formatted as plate-mjd-fiberid.fits & & bytes25 \\
sdss\_ra & SDSS Right Ascension (ICRS) & $\deg$ & float64 \\
sdss\_dec & SDSS Declination (ICRS) & $\deg$ & float64 \\
sdss\_class & SDSS spectroscopic class & & bytes32 \\
sdss\_redshift & SDSS spectroscopic redshift & & float32 \\ 
sdss\_redshift\_err & SDSS spectroscopic redshift uncertainty & & float32 \\ 
sdss\_redshift\_warn & SDSS spectroscopic redshift warning flag & & float32 \\ 
sdss\_u & Flux in the SDSS $u$ band & Jy & float32 \\
sdss\_u\_err & 1$\sigma$ flux uncertainty in the SDSS $u$ band & Jy & float32 \\
sdss\_g & Flux in the SDSS $g$ band & Jy & float32 \\
sdss\_g\_err & 1$\sigma$ flux uncertainty in the SDSS $g$ band & Jy & float32 \\
sdss\_r & Flux in the SDSS $r$ band & Jy & float32 \\
sdss\_r\_err & 1$\sigma$ flux uncertainty in the SDSS $r$ band & Jy & float32 \\
sdss\_i & Flux in the SDSS $i$ band & Jy & float32 \\
sdss\_i\_err & 1$\sigma$ flux uncertainty in the SDSS $i$ band & Jy & float32 \\
sdss\_z & Flux in the SDSS $z$ band & Jy & float32 \\
sdss\_z\_err & 1$\sigma$ flux uncertainty in the SDSS $z$ band & Jy & float32 \\
wise\_id & Unique WISE designation & & bytes19 \\
wise\_ra & WISE Right Ascension (ICRS) & $\deg$ & float64 \\
wise\_dec & WISE Declination (ICRS) & $\deg$ & float64 \\
wise\_W1 & Flux in the WISE \textit{W1} band & Jy & float64 \\
wise\_W1\_err & 1$\sigma$ flux uncertainty in the WISE \textit{W1} band & Jy & float64 \\
wise\_W2 & Flux in the WISE \textit{W2} band & Jy & float64 \\
wise\_W2\_err & 1$\sigma$ flux uncertainty in the WISE \textit{W2} band & Jy & float64 \\
wise\_W3 & Flux in the WISE \textit{W3} band & Jy & float64 \\
wise\_W3\_err & 1$\sigma$ flux uncertainty in the WISE \textit{W3} band & Jy & float64 \\
wise\_W4 & Flux in the WISE \textit{W4} band & Jy & float64 \\
wise\_W4\_err & 1$\sigma$ flux uncertainty in the WISE \textit{W4} band & Jy & float64 \\
\enddata
\tablecomments{Description of columns of our data table describing the WISE Matched SNR sample. Flux and flux uncertainty values not used in the analysis of the objects will be masked.}
\end{deluxetable*}

\startlongtable
\begin{deluxetable*}{llcc}
\tablecaption{Description of Photometric Results Table Columns\label{tab:photores}}
\tablehead{
    \colhead{Name} &
    \colhead{Description} &
    \colhead{Unit} &
    \colhead{Type}
}
\startdata
sdss\_id & Unique SDSS objID & & int64 \\
P\_AGN & Probability of AGN fit from F-Test & & float64 \\
E\_AGN & E template fit coeff. in the AGN fit & & float64 \\
E\_AGN\_err & $1\sigma$ E template fit coeff. uncertainty in the AGN fit & & float64 \\
IM\_AGN & IM template fit coeff. in the AGN fit & & float64 \\
IM\_AGN\_err & $1\sigma$ IM template fit coeff. uncertainty in the AGN fit & & float64 \\
Sbc\_AGN & Sbc template fit coeff. in the AGN fit & & float64 \\
Sbc\_AGN\_err & $1\sigma$ Sbc template fit coeff. uncertainty in the AGN fit & & float64 \\
AGN & AGN template fit coeff. & & float64 \\
AGN\_err & $1\sigma$ AGN template fit coeff. uncertainty & & float64 \\
EBV\_AGN & AGN E($B-V$) & & float64 \\
EBV\_AGN\_err & $1\sigma$ AGN E($B-V$) uncertainty & & float64 \\
logP\_AGN & Log posterior of the AGN fit ($-\chi^2$) & & float64 \\
logP\_AGN\_err & $1\sigma$ log posterior uncertainty of the AGN fit ($-\chi^2$) & & float64 \\
F\_AGN\_EBV & Integrated flux in the AGN component (E($B-V$) uncorrected) & Jy\,PHz & float64 \\
F\_AGN\_EBV\_err & $1\sigma$ integrated flux uncertainty in the AGN component (E($B-V$) uncorrected) & Jy\,PHz & float64 \\
F\_E\_AGN & Integrated flux in the E component for the AGN fit & Jy\,PHz & float64 \\
F\_E\_AGN\_err & $1\sigma$ integrated flux uncertainty in the E component of the AGN fit & Jy\,PHz & float64 \\
F\_IM\_AGN & Integrated flux in the IM component for the AGN fit & Jy\,PHz & float64 \\
F\_IM\_AGN\_err & $1\sigma$ integrated flux uncertainty in the IM component of the AGN fit & Jy\,PHz & float64 \\
F\_Sbc\_AGN & Integrated flux in the Sbc component for the AGN fit & Jy\,PHz & float64 \\
F\_Sbc\_AGN\_err & $1\sigma$ integrated flux uncertainty in the Sbc component of the AGN fit & Jy\,PHz & float64 \\
F\_AGN & Integrated flux in the AGN component (E($B-V$) corrected) & Jy\,PHz & float64 \\
F\_AGN\_err & $1\sigma$ integrated flux uncertainty in the AGN component (E($B-V$) corrected)  & Jy\,PHz & float64 \\
E\_Gal & E template fit coeff. in the Gal fit & & float64 \\
E\_Gal\_err & $1\sigma$ E template fit coeff. uncertainty in the Gal fit & & float64 \\
IM\_Gal & IM template fit coeff. in the Gal fit & & float64 \\
IM\_Gal\_err & $1\sigma$ IM template fit coeff. uncertainty in the Gal fit & & float64 \\
Sbc\_Gal & IM template fit coeff. in the Gal fit & & float64 \\
Sbc\_Gal\_err & $1\sigma$ Sbc template fit coeff. uncertainty in the Gal fit & & float64 \\
logP\_Gal & Log posterior of the Gal fit ($-\chi^2$) & & float64 \\
logP\_Gal\_err & $1\sigma$ log posterior uncertainty of the Gal fit ($-\chi^2$) & & float64 \\
F\_E\_Gal & Integrated flux in the Sbc component for the Gal fit & Jy\,PHz & float64 \\
F\_E\_Gal\_err & $1\sigma$ integrated flux uncertainty in the E component of the Gal fit & Jy\,PHz & float64 \\
F\_IM\_Gal & Integrated flux in the IM component for the Gal fit & Jy\,PHz & float64 \\
F\_IM\_Gal\_err & $1\sigma$ integrated flux uncertainty in the IM component of the Gal fit & Jy\,PHz & float64 \\
F\_Sbc\_Gal & Integrated flux in the Sbc component for the Gal fit & Jy\,PHz & float64 \\
F\_Sbc\_Gal\_err & $1\sigma$ integrated flux uncertainty in the Sbc component of the Gal fit & Jy\,PHz & float64 \\
\enddata
\tablecomments{Descriptions of columns of our Photometric Results Table. Fluxes are integrated from 0.1 to 30\,$\mu$m in the rest frame.}
\end{deluxetable*}

\bibliographystyle{aasjournal}
\bibliography{bibliography}{}

\end{document}